\pdfoutput=1

\documentclass[aps,showpacs,preprintnumbers,amsmath,amssymb,
eqsecnum, twocolumn, tightenlines
]{revtex4}

\usepackage{graphicx}

\newcommand{\be}{\begin{eqnarray}}
\newcommand{\ee}{\end{eqnarray}}

 \newcommand{\gsim}{\mathrel{\hbox{\rlap{\lower.55ex \hbox {$\sim$}}
                   \kern-.3em \raise.4ex \hbox{$>$}}}}
\newcommand{\lsim}{\mathrel{\hbox{\rlap{\lower.55ex \hbox {$\sim$}}
                   \kern-.3em \raise.4ex \hbox{$<$}}}}

\def\roughly#1{\mathrel{\raise.3ex\hbox{$#1$\kern-.75em%
\lower1ex\hbox{$\sim$}}}}
\def\lsim{\roughly<}
\def\gsim{\roughly>}

\newcommand{\ba}{\begin{eqnarray}}
\newcommand{\ea}{\end{eqnarray}}

\begin{document}
\title{  The  Fate of the Initial State Fluctuations in Heavy Ion Collisions. \\ III
The Second Act  of Hydrodynamics }
\author {Pilar Staig and Edward Shuryak}
\address { Department of Physics and Astronomy, State University of New York,
Stony Brook, NY 11794}
\date{\today}
\begin{abstract}
The hydrodynamical description of the ``Little Bang" in heavy ion collisions is surprisingly successful,  mostly due to the very small viscosity of the Quark-Gluon plasma. In this paper we
systematically study the propagation of small perturbations, also treated hydrodynamically. We start with a number of known
techniques allowing for the analytic calculation of the  propagation of small perturbations on top of the expanding fireball. The simplest approximation is the ``geometric acoustics", which substitutes the wave equation by mechanical equations for the propagating ``phonons". Next we turn to the case in which variables can be separated, where one can obtain not only the eikonal phases but also the amplitudes of the perturbation. Finally, we focus on the so called {\em Gubser flow}, a particular conformal analytic solution for the fireball expansion, on top of which one can derive closed equations for small perturbations. Perfect hydrodynamics allows all variables to be separated and all equations to be solved in terms of known special functions. We can thus collect the analytical expression for all the harmonics and reconstruct the complete Green function of the problem. In the viscous case the equations still allow for variable separation, but one of the equations has to be solved numerically. Summing all the harmonics we show real-time  perturbation evolution, observing the viscosity-induced changes in the spectra and the correlation functions. The calculated angular shape of the correlation function is remarkably similar to the shape emerging from the experimental data, for sufficiently large viscosity. We predict a minimum at $m\sim 7$ and maximum at $m\sim 9$ harmonics, which also have some experimental evidence for it. We conclude that local ``hot spots" in the initial state are the only visible origin of the observed correlations. 
\end{abstract}
\maketitle
\section{Introduction}
Since it is the third paper of the series devoted to the propagation of perturbations on top of the ``Little Bang", it does not need a detailed introduction. Let us only briefly point out the main physics of the phenomena in question, and then mention where the reader can find important earlier works on the subject.

Initial state perturbations of an ``average fireball", which occur on an event-by-event basis, lead to divergent sound waves, similar to the circles from a stone thrown into a pond. The sound velocity is $\sim 1/2$ and the time till freezeout $\tau_{FO}\sim 2R$ (where R is the nuclear size, about 6 fm for Au nuclei used in the experiment), thus the ``sound horizon" (the maximal radius of the circles) reaches $H_s\sim R$. In terms of the angular variables we use, it means a response at relatively large angles, $O(\pm 1 \, radian)$, from the perturbation. The strong radial explosion of the fireball dramatically enhances the contrast, making small deviations of the freezeout surface easily observable experimentally, provided the transverse momenta of the particles are tuned into the appropriate range. The shape of the hydro response to an initial point perturbation (the Green function) is quite non-trivial, and we show that for appropriate values of the viscosity it  reproduces  the shapes of the two-point correlation functions observed experimentally surprisingly well. We will conclude 
with the ``minimal" and ``maximally coherent" scenarios of the collisions: for experimental selection between those
one needs to measure certain three-point correlations functions, as was discussed in detail in our previous paper \cite{Staig:2010pn}.

Many issues we discuss, such as the power spectrum of higher harmonics of perturbations, are analogous to  the events in Cosmology during the last decade. We mean in particular the observations of the sound horizon scale, both in the   cosmic microwave background (CMB) radiation (see e.g.\cite{7WMAP} and the earlier work cited in it) and in the distribution of galaxies \cite{Eisenstein:2005su}. Discussing similarities and differences between the Little and Big Bangs  will be a recurring theme of this paper. Let us just comment that while these observations did turn Cosmology into a much more quantitative science, hopefully their ``Little Bang" analogues will also help us  to fix the global parameters of  nuclear collisions and the QGP much better.

Outlining the paper's context, we now go into a bit more detail
over the brief history of the ``second act of hydro". Sound
propagation on top of the expanding fireball  was first considered by Casalderrey-Solana and one of us (ES) in \cite{CasalderreySolana:2005rf}.  The fireball expansion was modelled by a Universe expansion using the Friedmann-Lemetre-Robertson-Walker metrics, and the specific phenomena discussed in it was the effect of the variable speed of sound (due to the QCD phase transition) on sound propagation. Its main result was the appearance of backward-moving or $convergent$ spherical/conical waves, together with the usual divergent ones. It is worth noting that the hadronic era has a near-constant speed of sound $cs \approx 0.4$ as noted in \cite{Shuryak:1972zq} and established later for the chemically non-equilibrated version of hadronic matter in \cite{Teaney:2002aj}. The so called mixed phase era is the only one in which $c_s$ varies. 

A qualitative picture of  the ``sound circles" resulting from point-like initial-state perturbations, and reaching by freezeout the so called ``sound horizon" radius, were introduced in the first paper of this series \cite{Shuryak:2009cy}. 

(It is amusing  that Gurzadyan and Penrose \cite{Gurzadyan:2010da} not only came  claim that the WMAP data provide some evidences for circles, or even co-central circles, in the CMB temperature variations. The ones they found, however, have sizes few times $larger$ than the sound horizon scale. So, if the claim is statistically sound, those must be some pre-Big-Bang events.)

Unlike the Big Bang, for which one reads the temperature perturbations  from the sky, the observable traces of the sound circles in the Little Bang are not so direct. The temperature and velocity perturbations both contribute to the particle spectra at the freezeout, and the picture is strongly affected by strong radial flow and the existence of the fireball's boundaries. The contribution of all of this to the spectra predicted in \cite{Shuryak:2009cy} was the ``double-horn" shape of the angular distribution, with two maxima identified with the latest crossings of the sound circle with  the fireball boundaries.  The ``circle" phenomenon has also been found by the Brazilian group, in their (zero viscosity)  numerical studies of ``event-by-event hydrodynamics" \cite{Andrade:2009em}. This group however went further and calculated the two-body correlators, finding  their characteristic three-maxima structure. The details of such structure in our (viscous) solution will be compared to the experimental data at the end of this
paper.

A  general setting of the problem, including the identification of the two basic scales of the problem, the so called  ``sound
horizon" and ``viscous horizon", was made in the second paper
of the series \cite{Staig:2010pn}, in which we also studied in
detail the perturbations using the geometric Glauber model.
Similar ideas have also been  proposed by Mocsy and Sorensen in
\cite{Mocsy:2010um,comment1}.

 The impetus for experimental studies of perturbation-related
effects  was provided by the paper by Alver and Roland
\cite{Alver:2010gr}.  They have pointed out that the two particle correlation data contains large $third$ angular harmonics, and attributed it to the  ``triangular" shapes of some events. That prompted many studies of the initial perturbations in the Glauber model, in which the fluctuations are due to the random positions and the interaction probability of the colliding nucleons inside the nuclei. It has been found  that the $v_3$ data can indeed be explained, by Glauber estimates of the initial perturbations $<\epsilon_3(in)^2>$ times the ``hydro response" at freezeout (fo) $\textrm{v}_3(fo)/\epsilon_3(in)$.    

In general, there are two  different views on the nature of the perturbations. A priori, the structure of the initial state perturbations can either (i)  be just Gaussian noise or (ii) contain important correlations between the harmonics. In the former case,  the ``minimal Gaussian scenario" of the initial state, the set of input parameters $<\epsilon_m(in)^2>$ has all the information one may possibly need, and all that needs to be done is the hydrodynamical calculation of  the ``linear response" ratios $\textrm{v}_n(p_t,fo)/\epsilon_n(in)$ between the initial perturbations of the fireball shape and the final flow, for each of the harmonics. The other school was pioneered by the above-mentioned Brazilian group, that started to do ``event-by-event hydrodynamics" for many (hundred thousands)  initial conditions, provided by certain event generator. Clearly this only makes sense if one hopes to reproduce  certain non-trivial correlations contained in experiment in a statistically significant way.

Our study in the previous work  \cite{Staig:2010pn} , based on Glauber theory,  had indeed found  non-trivial phase correlations between all $odd$ harmonics $m=1,3,5...$. We have ascribed those to the so called ``hot/cold spots" in the initial matter distributions, which can appear at any angle and are mutually uncorrelated.  We also pointed out the role of  the higher correlators and the ``resonance condition" between
three (or more) harmonics in order to measure the relative phases. Similar studies have also been done elsewhere, see e.g \cite{Teaney:2010vd}  focused on the resonance between the first and third harmonic with the second (the reaction plane) and the triangular flow.

Our main aim in this work is to derive the magnitude of all  harmonics of the flow in the same setting. Only then one can study their coherent sum, the Green function etc. This goal is achieved (semi) analytically, with separation of variables and full inclusion of viscosity effects. New important phenomenon -- existence of acoustic dips and peaks  in the power spectrum -- is suggested, calculated  and  correlated with experimental data.

However these breakthroughs came with a prize: we consider (i) only the central collisions, (ii) only conformal EoS of matter and (iii) only small perturbations.  The reader should be aware of the fact that our results should capture the qualitative behaviour rather than produce accurate numbers, directly related to the experimental data. Corrections to non-conformity and non-linearity as well as not-too-large non-centrality can be also studied, but those will be done elsewhere. 

Let us only comment here on the issue of non-linearity. If (as we believe) all harmonics add up coherently, the  perturbations are generally not small, $O(1)$, at the initial time. However,
as the perturbation expands and becomes a large sound circle (with the radius up to the ``sound horizon" size comparable to that of the fireball itself), it quickly becomes small. This is especially true for higher harmonics (to which this paper is mostly devoted) as they are additionally suppressed by viscosity.
  
  Clearly, early time evolution of perturbations is nonlinear, and releated effects are not captured by our approach. One practically important issue here is the speed of the waves, which affects the size of the sound circle at freezeout, which subsequently determines positions of the peaks in the correlation function and  the power spectrum.  Finite amplitude waves are known to travel faster than sound. We had investigated this correction and will include it in  our subsequent paper. Its effect is rather modest: for example a  factor 2 matter compression leads to only 15\% increase in speed.  Note, that realistic EoS leads to the speed of sound at late stages of the collision to be about 20\% lower than $c_s=1/\sqrt{3}$ in our conformal liquid:  these two effects to certain degree cancel each other.  

Let us further note, that at the initial time the pressure and flow gradients are especially large at local density fluctuations.  Therefore the applicability conditions of  even (viscous) hydrodynamics itself should be investigated. Interesting effects, such as e.g. cavitation, are known to occur in other hydro applications in similar settings. Theoretically, the issue is what the sum of all large gradients times
corresponding dissipative coefficients can actually do if resummed: see recent discussion in Ref. \cite{LS} on ``resummed hydrodynamics" and its applicability, for AdS/CFT and heavy ion collisions.

Returning to hydrodynamics, we would like to address the issues in the case with maximal symmetry: therefore we only discuss the central collisions, which are axially symmetric (without perturbations). We also would like to be as transparent as possible, thus using the analytic tools.  Finally, we believe that  in this problem, as in many others, one should look for  the Green function, the solution with an elementary delta-function-like source. Once it is found, any type of initial conditions  can be easily included by just a convolution with the Green function.  From the physics point of view it seems to be more important  to calculate the effect of the {\em viscosity} on the shape of the angular response, rather than to include the non-linear interactions between the harmonics, as ``event-by-event hydrodynamics" does.

The paper is organized as follows. We start by discussing two
approaches which can be used in the case when the perturbation
size is much smaller than the size of the system, so that the
number of excited harmonics is large. One  is the general ``geometric acoustics" method,   which substitutes the wave equation by mechanical equations for the propagating ``phonons". The other uses the standard eikonal representation of the solution, plus separation of variables. Finally, we focus on the so called Gubser flow, the conformal analytic solution for the fireball expansion \cite{Gubser:2010ze} with longitudinal and transverse flows. Significant further development  is due to Gubser and Yarom \cite{Gubser:2010ui}, who derived the linearized equations for  the propagation of small perturbations around it. In our paper we extend  their results to sound Green function, the coherent sum of all harmonics describing the propagating sound from a point-like ``hot spot". Our next step is to focus on  how the perturbations modify the freezeout surfaces and thus observed spectra and correlators, and finally compare the latter to the data.
\subsection{Relativistic   hydrodynamics, the zeroth-order}

By the ``zeroth order"  hydrodynamical evolution of the system we mean the one in which all possible perturbations of the ``average fireball shape" are not included. Additional simplifications often used are due to (approximate) symmetries which the problem possesses, for example rapidity-independence and also consideration of only central (axially symmetric) collisions. If those are assumed, the number of variables is reduced from 4 to 2, and one may start thinking about its analytic treatment. Otherwise, the problem only allows for numerical solutions, which are widely used in practice but will not be discussed in this work.

Our main goal, as we proceed, will be  to go to the ```first
approximation", deriving small perturbations of the zeroth-order
solution.  Unlike the zeroth order, the perturbations are not assumed to have any {\em a priori} symmetries. The main object of the hydrodynamical description, the stress tensor, is conserved:  thus the  equations to be solved are written as its zero covariant divergence %
\be T^{\mu\nu}_{;\mu} = \left(T^{\mu\nu}_{(0)}+ \delta T^{\mu\nu}_{(1)} \right)_{;\mu}=0\ee%
where the zero and one in parenthesis  are not the indices but
the order of perturbation. The perturbation term will be
assumed to be small and treated in the linear approximation.

While it is all very generic,  for completeness of the paper let
us  remind some details here, starting with the simplest example
of rapidity-independent ``Bjorken" flow. Even in this case, one
needs curved coordinates with a non-trivial metric, thus covariant derivatives and the nonzero Christoffel symbols will be needed:%
\begin{eqnarray}
{T^{ik}}_{;p} &=& {T^{ik}}_{,p} + \Gamma^i_{pm} T^{mk}
                                   + \Gamma^k_{pm} T^{im}\,,
                                   \label{eqn_Christ}
\end{eqnarray}
Changing Minkowski coordinates $t,x,y,z$, with $z$ along the beam, to the hyperbolic-cylindrical set $\tau,\eta,r,\phi$
 \be
   t = \tau \cosh\eta &     &   z = \tau \sinh\eta
 \\    x=r \cos\phi, &     &  y=r \sin\phi \\
  \tau = \sqrt{t^2-z^2} &     &  \eta = {1 \over 2} \ln{ \left( t{+}z\over t{-}z  \right)}
 \ee
one finds the following metric tensor
 \be
   g_{mn}=\left( \begin{array}{*{4}{c}}
        -1 & 0 & 0 & 0 \\
        0 & \tau^2 & 0 & 0 \\
        0 & 0 & 1 & 0 \\
        0 & 0 & 0 & r^2 \\
        \end{array} \right)\,,
 \ee
and the Christoffel symbols, following from standard expression
 \be
    \Gamma^s_{ij} =
    (1/2) g^{ks} \bigl( g_{ik,j} + g_{jk,i} - g_{ij,k}\bigr)\,.
 \ee
have the following non-vanishing components
 \be
   \Gamma^{\eta}_{\eta \tau} =
   \Gamma^{\eta}_{\tau \eta} = {1\over\tau}\,,\qquad
   \Gamma^\tau_{\eta \eta} = \tau \nonumber\\
\Gamma^{\phi}_{\phi r} =
   \Gamma^{\phi}_{r \phi} = {1\over r}\,,\qquad
   \Gamma^r_{\phi \phi} = -r
 \ee
Those are inserted into (\ref{eqn_Christ}) together with the general expression for relativistic Navier-Stokes stress tensor
\begin{eqnarray}
T_{\mu\nu} = (\epsilon + p) u_{\mu}u_{\nu}+pg_{\mu\nu}-2\eta\sigma_{\mu\nu}-
\zeta(u^{\lambda}_{\phantom{a} ;\lambda})\Delta_{\mu\nu},
\end{eqnarray}
where,
\begin{eqnarray}
\sigma_{\mu\nu} & = & \Delta_{\mu}^{\alpha} \Delta_{\nu}^{\beta}\left(\frac{u_{\beta;\alpha} +u_{\alpha;\beta}}{2}-\frac{g_{\alpha\beta}}{3} u^{\lambda}_{;\lambda}\right)\\
\Delta_{\mu\nu} & = & u_{\mu}u_{\nu}+g_{\mu\nu}
\end{eqnarray}
The first two terms of the stress-energy tensor correspond to ``ideal hydrodynamics", while  the third and fourth ones are due to shear and bulk
viscosity, respectively.
 
The corresponding analytic solution, known as  the Bjorken flow,
\cite{Bjorken:1982qr} corresponds to colliding objects being infinite walls of matter, eliminating the transverse flow and any dependence on the two transverse coordinates $x,y$ or $r,\phi$, as well as on $\eta$. Furthermore, we will consider the simplest {\em co-moving flow} case, with a trivial 4-velocity $u^\mu=(1,0,0,0)$. Then the non-viscous stress tensor returns to its  generic form  
\be
T^{\mu\nu}=diag(\epsilon(\tau), p(\tau),p(\tau),p(\tau)) \ee 
in the medium rest frame, depending on the proper time $\tau$. The resulting 00 and 11 equations, together with the thermodynamic identity relating the differentials of these quantities
   \be
     {\partial_\mu \epsilon \over \epsilon+p}= {\partial_\mu s \over s}
  \ee
can be put  into the final form of one single ``entropy production equation"
  \be
{ds\over d\tau}={s \over\epsilon+p} {d\epsilon \over d\tau} =-{s \over
   \tau}\left(1-{(4/3)\eta+\xi \over (\epsilon+p)  \tau}\right)
      \ee
Note that if both viscosities are zero, the solution is just
$s\tau=const$, which implies simply the  total entropy
conservation.
\section{Sound propagation in the short-wavelength approximation}
\subsection{The geometric acoustics}
If the wavelength of the perturbation is small compared to the
size of the system,  one can describe sound propagation in the
``geometric acoustics" approximation, see textbooks such as
\cite{LL_fluid}. The reason we can use such an approximation in our problem is the assumed $locality$ of the initial ``hot spots"  (and thus the initial width of the propagating circular wave). All we need is that their size is much smaller than the fireball dimensions \be l \ll R \ee

 The derivation of the approximation is based on the analogy between the Hamilton-Jacobi equation for the particle propagation and the wave  equation for the sound, deriving the Hamilton equations of motion for the ``sound particles" (``phonons" ).  The resulting equations of motion for them are
\be %
{d\vec{r} \over dt}= {\partial \omega(\vec{k},\vec{r}) \over \partial \vec{k}}  \label{eqn_speed}\,,  \\
{d\vec{k} \over dt}=-{\partial \omega(\vec{k},\vec{r}) \over
\partial \vec{r}}\,, \label{eqn_force}
\ee %
driven by the (position dependent) dispersion relation
$\omega(\vec{k},\vec{r})$.

 Let us start with the simplest non-relativistic case,  with  small velocity of the flow, $u \ll 1$. In this case the dispersion relation is obtained from that in the fluid at rest by a local Galilean transformation, so that for flow $\vec{u}(\vec r)$
\be %
\omega(\vec{k},\vec{r})=c_{s} k + (\vec{k}\,  \vec{u}(\vec r))\,.
\label{eqn_nonrel}
\ee %
As a simple yet relevant example, let us use the (generalized) Hubble flow in which the velocity profile is linear
\be %
u^{i}(r)=H^{ij} r^{j}\,,
\ee %
with some constant (time and coordinate independent) Hubble tensor. The eqn (\ref{eqn_force}) now reads as ``rotation" of the phonon momentum
\be %
{dk^{i} \over dt} = - H^{ij} k^{j}\,.
\ee %
If the Hubble tensor is symmetric, it can be diagonalized with
3 real eigenvalues, $H_{1},H_{2},H_{3}$, so the general solution in its eigenframe is the  exponential change of the corresponding momentum components $k_{i}(t)=exp(-H_{i}t) k_{i}(0)$. Note that if all three eigenvalues are the same, the unit vector of the direction $\vec{n}_{\vec{k}}$ would be time-independent. Furthermore, if the Hubble tensor contains an anti-symmetric part, the direction vector would be rotating around the vector $\epsilon_{ijk}H_{jk}$.

Let us now come to the first eqn (\ref{eqn_speed})
\be %
{dr^{i} \over dt} = c_{s} {n}^i_{\vec{k}}(t) + H^{ij}
r^{j}(t)\,.
\ee %
with the first term in the r.h.s. containing a unit vector along $\vec k$. The simplest case is when the Hubble matrix is proportional to the unit matrix and the first term is time-independent: then the solution is simply a linear addition of the sound motion and the Hubble expansion
\be %
\vec{r}(t)= c_{s} t \vec{n}_{\vec{k}} + \vec{r}(0) exp(+Ht)\,.
\ee %
This approximation is enough to explain the deformations which the zeroth-order flow induces on the basic geometric shapes of the sound fronts -- the cylinders, spheres or cones -- appearing in a non-floating medium. (We will use it for this purpose elsewhere \cite{withVlad}.) It is however not so useful for predicting the corresponding $amplitudes$ of the wave, which we will discuss in the next subsection.

\subsection{Wave equations with separable variables}
Let us explain the idea in the simplest setting, assuming that
there are only time and one relevant space coordinate, $x$.  Let us also assume that one can eliminate the velocity and write the hydrodynamic equations as a closed  second-order linear equation for the temperature perturbation $\delta(t,x)$

\be  {\partial^2 \delta \over \partial t^2}-  C_1(t,x)  {\partial^2 \delta \over \partial x^2}
+C_2 (t,x)  {\partial \delta \over \partial t}  \nonumber \\
+C_3 (t,x)  {\partial \delta \over \partial x}
+C_4 (t,x)  \delta=0
\ee
where $C_1..C_4$ are some functions.

The idea is similar to the semiclassical approximation in quantum mechanics, which uses for the wave function a form $\psi(t,x)\sim A(t,x)exp(iF(t,x)/\hbar) $, with some amplitude and the phase, assuming that the phase is parametrically large $F/\hbar \gg 1$. If so,  one can find a solution satisfying subsequently parts of the equation of the same magnitude.

Let us show how it works for the generic 2-d equation at hand. One also introduces the amplitude and the phase 
\be \delta(t,x)\sim
A(t,x)exp(i\phi(t,x)/\epsilon)  \ee 
with the $\hbar$ substituted by a dimensionless abstract small parameter $\epsilon$. Its substitution into the equation above yields three types of terms
 $${1\over \epsilon^2}[-\dot\phi^2+C_1 (\phi')^2) ]  $$
 $$+ {i\over \epsilon}[2 {\dot{A}\dot{\phi} \over A} +\ddot{\phi}  + C_2  \dot{\phi}  -2C_1{A' \phi' \over A} -C_1\phi''
+C_3 \phi']$$ 
\be + [{\ddot{A}\over A}-C_1 {A'' \over A} +C_2 {
\dot{A}\over A}  +C_3 {A'\over A}+ C_4 ]=0 \ee 
For small $\epsilon$ one starts from the first square bracket. If the first coefficient can be factored into functions of both variables, $C_1=C_{1t}(t)C_{1x}(x)$ it can readily be solved yielding 
\be
\phi(t,x)=k \left( \int^t \sqrt{C_{1t}(t_1)} dt_1\pm
\int^{x}{dx_1\over \sqrt{C_{1x}(x_1)} }\right) \ee 
where the separation of variables constant $k$, the ``wave vector", is assumed to be large. When $C_{1x}=1 \sqrt{C_{1t}}=c_s=const$ we have a function of $x-c_s t$, the usual propagating wave.

The amplitude  $A$ should be found from  the second approximation, the terms of the order ${1/\epsilon}$.  One may again get an explicit solution assuming the variables can be separated. Looking for the amplitude in a  factorizable form $A=A_{t}(t)A_{x}(x)$ one can see that the first three terms can be only dependent on $t$, provided $C_2$ depends on time only. The last three $O({1/\epsilon})$ terms  would be factorizable into $C_{1t}(t)$ times a function of $x$ if $C_3=C_{1t}(t)*C_{3x}(x)$. If so, the solution for both parts of the amplitudes are
$$
A_t(t) = exp\int_0^t  dt_1 [
\alpha \sqrt{C_{1t}(t_1)}-{\dot{C}_{1t}(t_1) \over 4C_{1t}(t1)}-C_{2t}(t_1)/2  ] $$ 
\be
A_x(x)=exp\int_0^x dx_1 [- {\alpha \over \sqrt{C_{1x}(x_1)}} +{C'_{1x}(x_1)) \over 4C_{1x}(x_1)} \nonumber \\
+{C_{3x}(x_1)) \over 2C_{1x}(x_1)} ] \ee%
A new separation-of-variable constant $\alpha$ formally appears
here, but it does not generate anything new in respect to what was already included in  the phase, so it can safely be put to zero.

Familiar examples of waves are e.g. the spherical and conical
waves, in which case the variables can be separated . Indeed, when the spatial part of the equation is d-dimensional Laplacian, one has  
\be C_1={1\over c_s^2},  C_2=0,  C_3={d-1 \over x} {1\over c_s^2} \ee and the corresponding amplitude decays with distance as \be A \sim {1\over x^{{d-1 \over 2} } } \ee
which is well known for spherical (d=3) and cylindrical
( d=2 ) waves.

As the reader will see later, the sound on top of Gubser's flow
can also be shown to have an amplitude depending on new variables $\rho,\theta$ in a factorizable way, which was not the case in the original coordinates, the proper time $\tau$ and $r$. Therefore, without introduction of these coordinates, one would not be able to solve the equation for the amplitude in such a simple factorized form.
\section{Perturbations on top of the Gubser flow }
\subsection{Summary of the Gubser flow}
The Gubser flow \cite{Gubser:2010ze,Gubser:2010ui} is a solution which keeps the boost-invariance and the
axial symmetry in the transverse plane of the Bjorken flow, but replaces the
translational invariance in the transverse plane
 by symmetry under a special conformal transformation.
Therefore, the matter is required to be conformal, with the EOS
\be \epsilon=3p \sim T^4 \ee and the speed of sound
$c_s=1/\sqrt{3}$. The solution has one dimensional parameter $q$
 via which the finite size of
the nuclei is introduced.

Working in the $(\tau,\eta,r,\phi)$ coordinates with the metric
\begin{eqnarray}
ds^2 & = & -d\tau^2 + \tau^2 d\eta^2 + dr^2 +r^2d\phi^2,
\end{eqnarray}
and assuming no dependence on the rapidity $\eta$ and azimuthal
angle $\phi$,  the 4-velocity  can be parameterized by only one function
\begin{eqnarray}
u_{\mu} &  = &
\left(-\cosh{\kappa(\tau,r)},0,\sinh{\kappa(\tau,r)},0\right)
\end{eqnarray}

Omitting the details from \cite{Gubser:2010ze}, the solution for the velocity and the energy density is
\begin{eqnarray}
v_\perp & = & \tanh{\kappa(\tau,r)}  =  \left(\frac{2q^2\tau
r}{1+q^2\tau^2 + q^2r^2}\right)
\end{eqnarray}
\begin{eqnarray}
\epsilon & = & \frac{\hat{\epsilon}_0 (2
q)^{8/3}}{\tau^{4/3}\left(1+2q^2(\tau^2 +
r^2)+q^4(\tau^2-r^2)^2\right)^{4/3}}
\end{eqnarray}
where $\hat{\epsilon}_0 $ is some normalization parameter.

In \cite{Gubser:2010ui} Gubser and Yarom re-derived the same
solution by going into the co-moving frame. In order to do so they rescaled the metric 
\begin{eqnarray}
ds^2 & = & \tau^2 d\hat{s}^2
\end{eqnarray}
and performed  a coordinate transformation from the $\tau,r$ to a new set $\rho,\theta$ given by:
\begin{eqnarray}
\sinh{\rho} & = & -\frac{1-q^2\tau^2+q^2r^2}{2q\tau}\label{rho_coord}\\
\tan{\theta} & = &
\frac{2qr}{1+q^2\tau^2-q^2r^2}\label{theta_coord}
\end{eqnarray}

In the new coordinates the rescaled metric reads:
\begin{eqnarray}
d\hat{s}^2 & = &-d\rho^2 + \cosh^2{\rho}\left(d\theta^2 +
\sin^2{\theta}d\phi^2\right)+d\eta^2
\end{eqnarray}
and we will use $\rho$ as the ``new time" coordinate and $\theta$ as a new ``radial" coordinate. In the new coordinates the fluid is at rest, so the velocity field has only nonzero $u_{\rho}$.

The relation between the velocity  in Minkowski space in the
$(\tau, r,\phi,\eta)$ coordinates and the one in the rescaled
metric in $(\rho,\theta,\phi,\eta)$ coordinates corresponds to:
\begin{eqnarray}
u_{\mu} & = & \tau \frac{\partial \hat{x}^{\nu}}{\partial
\hat{x}^{\mu}}\hat{u}_{\nu}\, , \label{u_transf}
\end{eqnarray}
while the  energy density transforms as:
$\epsilon=\tau^{-4}\hat{\epsilon}$.

The  temperature (in the rescaled frame, $\hat{T}=\tau f_*^{1/4}T$, with $f_*=\epsilon/T^4=11$ as in \cite{Gubser:2010ze}) is now dependent only on the new time $\rho$, and in the case with  nonzero viscosity the solution is
\footnotesize
\begin{eqnarray}
\hat{T} & = &\frac{\hat{T}_0}{(\cosh{\rho})^{2/3}} +\frac{H_0
\sinh^3{\rho}}{9 (\cosh{\rho})^{2/3}} \,
_2F_1\left(\frac{3}{2},\frac{7}{6};\frac{5}{2},-\sinh^2{\rho}\right)\nonumber\\
\label{back_T}
\end{eqnarray}\normalsize
where $H_0$ is a dimensionless constant made out of the shear
viscosity and the temperature, $\eta = H_0 T^{3}$ and $_2F_1$ is
the hypergeometric function. In the inviscid case the solution is just the first term of expression (\ref{back_T}), and of course it also conserves the entropy in this case. The picture of the explosion is obtained by transforming  this expression back to the $\tau,r$ coordinates and performing the appropriate rescaling.

\subsection{Perturbations of the Gubser flow}
Small  perturbations to the Gubser flow obey linearized equations which have also been derived in \cite{Gubser:2010ui}. We start with  the zero viscosity case, so that the background temperature (now to be called $T_b$) will be given by just the first term in (\ref{back_T}). The perturbations over the previous solution are defined by
\begin{eqnarray}
\hat{T} & = &  \hat{T}_b(1+\delta)\label{Tpertb}\\
\hat{u}_{\mu} & = & \hat{u}_{0 \,\mu} + \hat{u}_{1\mu}\label{upert}
\end{eqnarray}
with
\begin{eqnarray}
\hat{u}_{0 \,\mu} & = & (-1,0,0,0)\\
\hat{u}_{1\mu} & = & (0,u_{\theta}(\rho,\theta,\phi),u_{\phi}(\rho,\theta,\phi),0)\\
\delta & = & \delta(\rho,\theta,\phi)
\end{eqnarray}
The careful reader will notice here, that although  general
perturbations should not have any symmetries of the zeroth
solution, we have not listed rapidity among the variables. Indeed, we only consider  the perturbations which are
rapidity-independent. The reason for that is that the initial
state perturbations are initiated in the transverse plane but
rapidity-independent, so that the waves they  induce also
propagate in the transverse plane only.

Plugging expressions (\ref{Tpertb}),(\ref{upert}) into the
hydrodynamic equations and only keeping linear terms in the
perturbation, one can get a system of coupled 1-st order
differential equations. Furthermore, if one ignores the viscosity terms, one may exclude velocity and get the following (second order) closed equation for the temperature perturbation:
\begin{eqnarray}
& &\frac{\partial^2 \delta}{\partial \rho^2} -
\frac{1}{3\cosh^2{\rho}} \left( \frac{\partial^2 \delta}{\partial
\theta^2}  +\frac{1}{\tan{\theta}}\frac{\partial \delta}{\partial
\theta}+ \frac{1}{\sin^2{\theta}}\frac{\partial^2 \delta}{\partial
\phi^2}
\right)  \nonumber\\
& & +\frac{4}{3}\tanh{\rho}\frac{\partial \delta}{\partial \rho}=0
\label{T_pert_eqn}
\end{eqnarray}
As we will show, it has a number of remarkable properties.

\subsection{The short-wavelength approximation for the sound waves on top of the Gubser flow}
Before we proceed to the exact solution of this equation, let us
follow the procedure described in section IIB and study the
solution to equation (\ref{T_pert_eqn}) in the short wavelength
approximation. We start by looking for a factorized solution of
the form:
\begin{eqnarray}
\delta & = & e^{i(f_{\rho}(\rho) -
f_{\theta}(\theta)-f_{\phi}(\phi))}F_{\rho}(\rho)F_{\theta}(\theta)F_{\phi}(\phi)
\end{eqnarray}
where $f_i>>1$, such that the derivatives taken over the
exponential are dominant.  In this way, we study the equation
separating it in different equations depending on which power of
the derivatives over the exponent they have. The first step is to look only at the second derivatives because, since  they produce terms of second order in the exponent, they are the leading ones.
In this way we find:
\begin{eqnarray}
f_{\rho}(\rho) & = & \pm \frac{2}{\sqrt{3}}k\arctan{e^{\rho}} +
A\\
f_{\theta}(\theta) & = & \pm \int d\theta\sqrt{k^2-\frac{m^2}{\sin^2{\theta}}}+B\label{f_th}\\
f_{\phi}(\phi) & = & \pm m\phi +C
\end{eqnarray}
The integral in (\ref{f_th}) can be solved, but it gives a
cumbersome result. So in what follows (of this section) we will assume no $\phi$ dependence just to get an idea of the result. When we do this, the functions in the exponent reduce to:
\begin{eqnarray}
f_{\rho}(\rho) & = & \pm \frac{2}{\sqrt{3}}k\arctan{e^{\rho}} +
A\\
f_{\theta}(\theta) & = & \pm k\theta +B\\
\end{eqnarray}

The function $f_{\rho}(\rho)$ is almost linear in $\rho$ in the
region that we are interested in studying ($-2\lesssim \rho
\lesssim 1$), so we find the phase of the solution to be $\sim k\rho$ which means that we indeed expect to find solutions in the form of the sound wave propagation (in this region).

Now that we have found the functions in the exponent we look for
the wave amplitude by cancelling among themselves the terms with
the first power of the large exponent:  by doing this we find
the amplitudes to be
\begin{eqnarray}
F_{\rho}(\rho) & \sim &
\frac{1}{(\cosh{\rho})^{1/6}} \label{Frho}\\
F_{\theta}(\theta) & \sim &  \frac{1}{\sqrt{\sin{\rho}}}
\label{Ftheta}
\end{eqnarray}

\subsection{The exact separation of variables for the perturbation}
We have seen that in the short wavelength approximation we found a separable wave-like solution to equation (\ref{T_pert_eqn}), and now we would like to see if the exact solution can be found by using variable separation
$\delta(\rho,\theta,\phi)=R(\rho)\Theta(\theta)\Phi(\theta)$. It is indeed so. In the non-viscous case, that we are now discussing, each of the three equations
\begin{eqnarray}
R''(\rho)+\frac{4}{3}\tanh{\rho}R'(\rho)+\frac{\lambda}{3\cosh^2{\rho}}R(\rho)=0\\
\Theta''(\theta)+\frac{1}{\tan{\theta}}\Theta'(\theta)+\left(\lambda-\frac{m^2}{\sin^2{\theta}}\right)\Theta(\theta)=0\\
\Phi'(\phi)+m^2\Phi(\phi)=0
\end{eqnarray}
is analytically solvable, with the result \small
\begin{eqnarray}
R(\rho) & = & \frac{C_1 P_{-\frac{1}{2} +
\frac{1}{6}\sqrt{12\lambda+1}}^{2/3}(\tanh{\rho}) + C_2
Q_{-\frac{1}{2} +
\frac{1}{6}\sqrt{12\lambda+1}}^{2/3}(\tanh{\rho})}{(\cosh{\rho})^{2/3}}\nonumber\\
\Theta(\theta) & = & C_3P_l^m(\cos{\theta})+C_4Q_l^m(\cos{\theta})\nonumber\\
\Phi(\phi) & = & C_5 e^{im\phi} + C_6 e^{-im\phi}\label{exact}
\end{eqnarray}

\noindent\normalsize where $\lambda=l(l+1)$ and P and Q are
associated Legendre polynomials.  The part of the solution
depending on $\theta$ and $\phi$ can be combined in order to form spherical harmonics $Y_{lm}(\theta, \phi)$, such that
$\delta(\rho,\theta,\phi)\propto R_l(\rho)Y_{lm}(\theta,\phi)$.
This property should have been anticipated, as one of the main
ideas of Gubser has been to introduce a coordinate which together with $\phi$ make a map on a 2-d sphere.

The implications of that for the physics we are going to discuss
are as follows. While we will project the spectra and correlation function to the azimuthal angle $\phi$ and its Fourier components, we will be focussing on the quantum number $m$ conjugated to it. In particular, the community is very much focused on the ``triangular flow" with $m=3$. In principle, however, this is produced by many $l$- harmonics, providing the obvious condition $l \geq m$ holds. Harmonics with different $l$ have obviously different radial dependence. (We mention this point, because there was some controversy about the powers of $r$, especially for various definitions of the ``dipole flows" with $m=1$.)

Let us explore  the asymptotic behavior of the Legendre functions when $l>>1$ that is given by \cite{Gradshtein}:
\begin{eqnarray}
P_l^m(\cos{\theta}) & = &
\frac{2}{\sqrt{\pi}}\frac{\Gamma(l+m+1)}{\Gamma(l+3/2)}
\frac{\cos{((l+1/2)\theta-\frac{\pi}{4}+\frac{m\pi}{2})}}{\sqrt{2\sin{\theta}}}\nonumber\\
Q_l^m(\cos{\theta}) & = &
\sqrt{\pi}\frac{\Gamma(l+m+1)}{\Gamma(l+3/2)}
\frac{\cos{((l+1/2)\theta+\frac{\pi}{4}+\frac{m\pi}{2})}}{\sqrt{2\sin{\theta}}}\nonumber\\
\end{eqnarray}
\normalsize These expressions show that for large $l$ the solution presents oscillatory behavior in $\theta$ with an amplitude given by $\frac{1}{\sqrt{\sin{\theta}}}$. It is gratifying to see, that this is the same that we obtained in the short-wavelength approximation for $F_{\theta}(\theta)$ (eq.\ref{Ftheta}) in the previous section.

Now let us look into the $\rho$-dependent part of the solution in the large $l$ limit we have that the Legendre polynomials as a function of $\tanh{\rho}$ correspond to: \scriptsize
\begin{eqnarray}
P_l^m(\tanh{\rho}) & = &
\sqrt{\frac{2}{\pi}}\frac{\Gamma(l+m+1)}{\Gamma(l+3/2)} \sqrt{\cosh{\rho}}  \nonumber\\
{} & &\cos{\left(\left(l+\frac{1}{2}\right)\arccos{(\tanh{\rho})}-\frac{\pi}{4}+\frac{m\pi}{2}\right)}\nonumber\\
Q_l^m(\tanh{\rho}) & = &
\sqrt{\frac{\pi}{2}}\frac{\Gamma(l+m+1)}{\Gamma(l+3/2)}\sqrt{\cosh{\rho}}    \nonumber\\
{} & &
\cos{\left(\left(l+\frac{1}{2}\right)\arccos{(\tanh{\rho})}+\frac{\pi}{4}+\frac{m\pi}{2}\right)}\nonumber\\\label{Leg_as_tanh}
\end{eqnarray}\normalsize
Again we see an oscillatory behavior and a wave amplitude.  In
this case the amplitude is given by $\sqrt{\cosh{\rho}}$ and if we divide this by $(\cosh{\rho})^{2/3}$ as we have in the exact
solution (\ref{exact}) we get an amplitude for the wave of
$\frac{1}{(\cosh{\rho})^{1/6}}$, which is the same as we got in
the preceding section (\ref{Frho}) using the short wavelength
approximation.

So we have checked that for large $l$ $\delta(\rho,\theta,\phi)$, and therefore the temperature perturbation in the rescaled frame, $\hat{T}_1(\rho,\theta,\phi)=\hat{T}_b(\rho) \delta(\rho,\theta,\phi)$, does behave
like a sound wave.

\subsection{Propagation of the local initial-state perturbation}
Let us  study the propagation of the hydrodynamical response
induced by an initial perturbation on top of the background at some initial ``time" $\rho_0$, given by a Gaussian-shaped initial ``hot spot":
\begin{eqnarray}
\hat{T}_1(\rho_0,\theta,\phi) \propto e^{-\frac{\theta^2 +
\theta_0^2 - 2 \theta \theta_0 \cos{(\phi-\phi_0)}}{2
s^2}}\label{IC1}
\end{eqnarray}
We further assume that at the initial time there is no flow (momentum), only extra energy, so another initial condition is:
\begin{eqnarray}
\hat{u}_{\theta}(\rho_0)& = & 0\nonumber\\
\hat{u}_{\phi}(\rho_0) & = & 0
\end{eqnarray}
which define the initial derivative of the temperature perturbation, since \cite{Gubser:2010ui}
\begin{eqnarray}
\hat{u}_{l \,i}& = & v_l(\rho)\partial_i Y_{lm}(\theta,\phi)\nonumber\\
v_l(\rho) & = & \frac{3\cosh^2{\rho}}{l(l+1)}\frac{d
\delta_{l}}{d\rho} \label{vl}
\end{eqnarray}
where $i=\theta,\phi$. Thus we require
\begin{eqnarray}
\frac{\partial \delta_l}{\partial \rho}|_{\rho=\rho_0} & = & 0
\label{IC2}
\end{eqnarray}
The general solution for linear perturbations is
\begin{eqnarray}
\hat{T}_1(\rho,\theta,\phi) & = & \sum_l \sum_{m=-l}^{m=l}
 c_{lm} R_l(\rho)Y_{lm}(\theta,\phi)\label{Tpert}\\ 
\hat{u}_{1 \, i}(\rho,\theta,\phi) & = & \sum_l \sum_{m=-l}^{m=l}
 c_{lm} v_l(\rho)\partial_i Y_{lm}(\theta,\phi) \label{vpert}
\end{eqnarray}
with \scriptsize
\begin{eqnarray}
R_l(\rho) & = &\frac{A_l
P_{-\frac{1}{2}+\frac{1}{6}\sqrt{12l(l+1)+1}}^{2/3}(\tanh{\rho}) +
B_l
Q_{-\frac{1}{2}+\frac{1}{6}\sqrt{12l(l+1)+1}}^{2/3}(\tanh{\rho})}{(\cosh{\rho})^{4/3}}\nonumber\\
\end{eqnarray}\normalsize
where $c_{lm}$, $A_l$ and $B_l$ are constants that can be
determined using the initial conditions (\ref{IC1}) and
(\ref{IC2}).  With $A_l$ and $B_l$ determined, the
$\rho$-dependent part of the temperature is
\begin{eqnarray}
R_l(\rho) & = &
\left(\frac{\cosh{\rho_0}}{\cosh{\rho}}\right)^{2/3}\delta_l(\rho) \label{Rl}\\
\delta_l(\rho) & = &\frac{\frac{d q_l}{d \rho}|_{\rho_0}p_l(\rho)
- \frac{d p_l}{d \rho}|_{\rho_0} q_l(\rho)}{\frac{d q_l}{d
\rho}|_{\rho_0}p_l(\rho_0) - \frac{d
p_l}{d \rho}|_{\rho_0} q_l(\rho_0)}\nonumber\\
\label{deltaI}
\end{eqnarray}
with
\begin{eqnarray}
p_l(\rho) & = &
\frac{P_{-\frac{1}{2}+\frac{1}{6}\sqrt{12l(l+1)+1}}^{2/3}(\tanh{\rho})}{(\cosh{\rho})^{2/3}}\\
q_l(\rho) & = &
\frac{Q_{-\frac{1}{2}+\frac{1}{6}\sqrt{12l(l+1)+1}}^{2/3}(\tanh{\rho})}{(\cosh{\rho})^{2/3}}
\end{eqnarray}
which, together with the $Y_{lm}(\theta,\phi)$, finally provides the complete solution for $\delta$ in equation (\ref{T_pert_eqn}) .  The denominator of the right term of $R_l(\rho)$ is the so called Wronskian of the functions $p_l(\rho)$ and $q_l(\rho)$ evaluated at the initial ``time" $\rho_0$. Since the Legendre polynomials P and Q are
linearly independent, the Wronskian is always non-zero, so we are guaranteed that the function $R_l(\rho)$ is always finite.

\begin{figure}[!h]
\begin{center}
\includegraphics[width=8 cm]{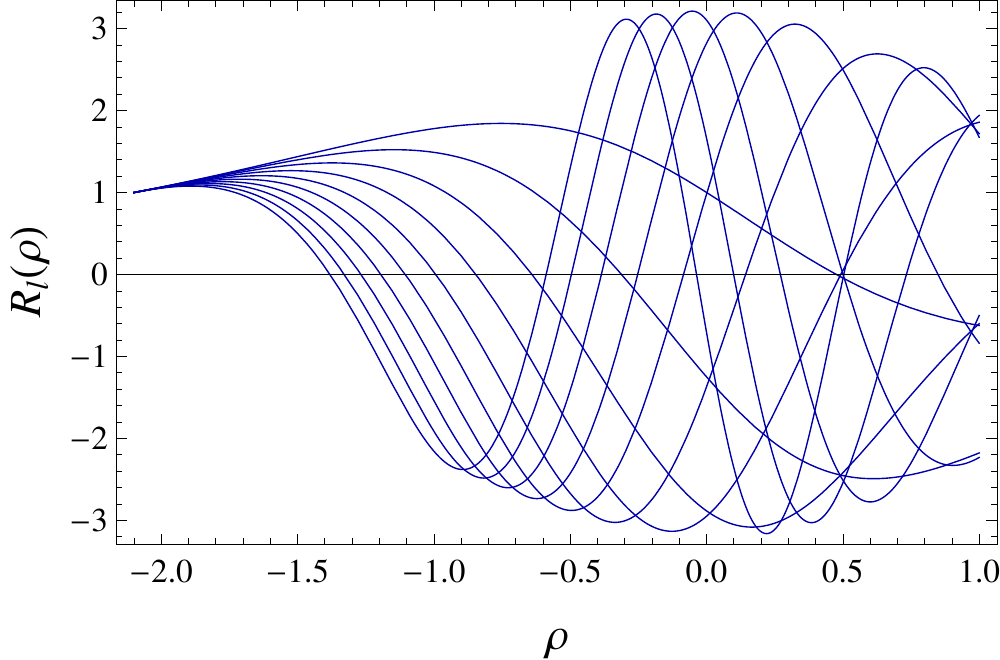}
\includegraphics[width=8 cm]{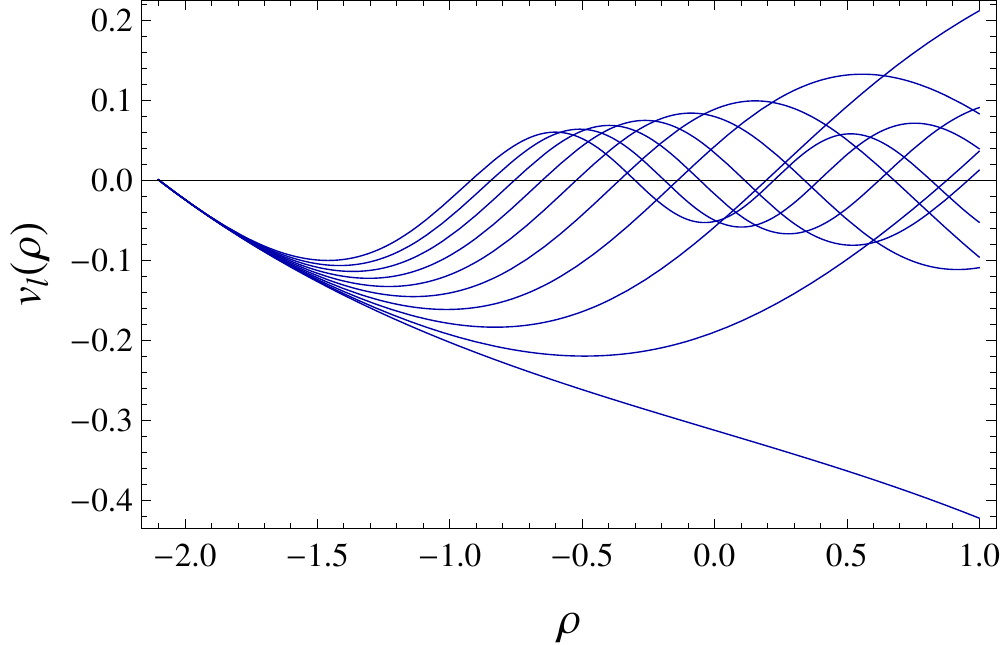}
\end{center}
\vspace{-5ex}\caption{(Color online) Top: $R_l(\rho)$ defined in (\ref{Rl}) for harmonic number $l$ from 1 to 10 (from less to more oscillating ones).
All curves are arbitrarily normalized to $1$ at $\rho=-2.07$.
Bottom: the corresponding harmonics of the velocity $v_l(\rho)$, defined in (\ref{vl}),  also for $l$ from 1 to
10. Note its oscillatory behavior for larger $l$ and later times.
}\label{amplitudes}
\end{figure}

The first ten harmonics  for $R_l$ and $v_l$ are plotted in
Fig.\ref{amplitudes}, showing how the amplitude varies as a
function of ``time" $\rho$. One can see how the initial
deformation (the upper plot, set to one for each l for comparison) is transferred into the flow velocity (the lower plot). One can also see that while for the lower harmonics it happens in a more or less linear way, higher harmonics show oscillating behaviour , as expected for sound waves. Indeed, one should see a transition from potential to kinetic energy happening with higher and higher rate, as the harmonic number grows.

\begin{figure}[!h]
\begin{center}
\includegraphics[width=8 cm]{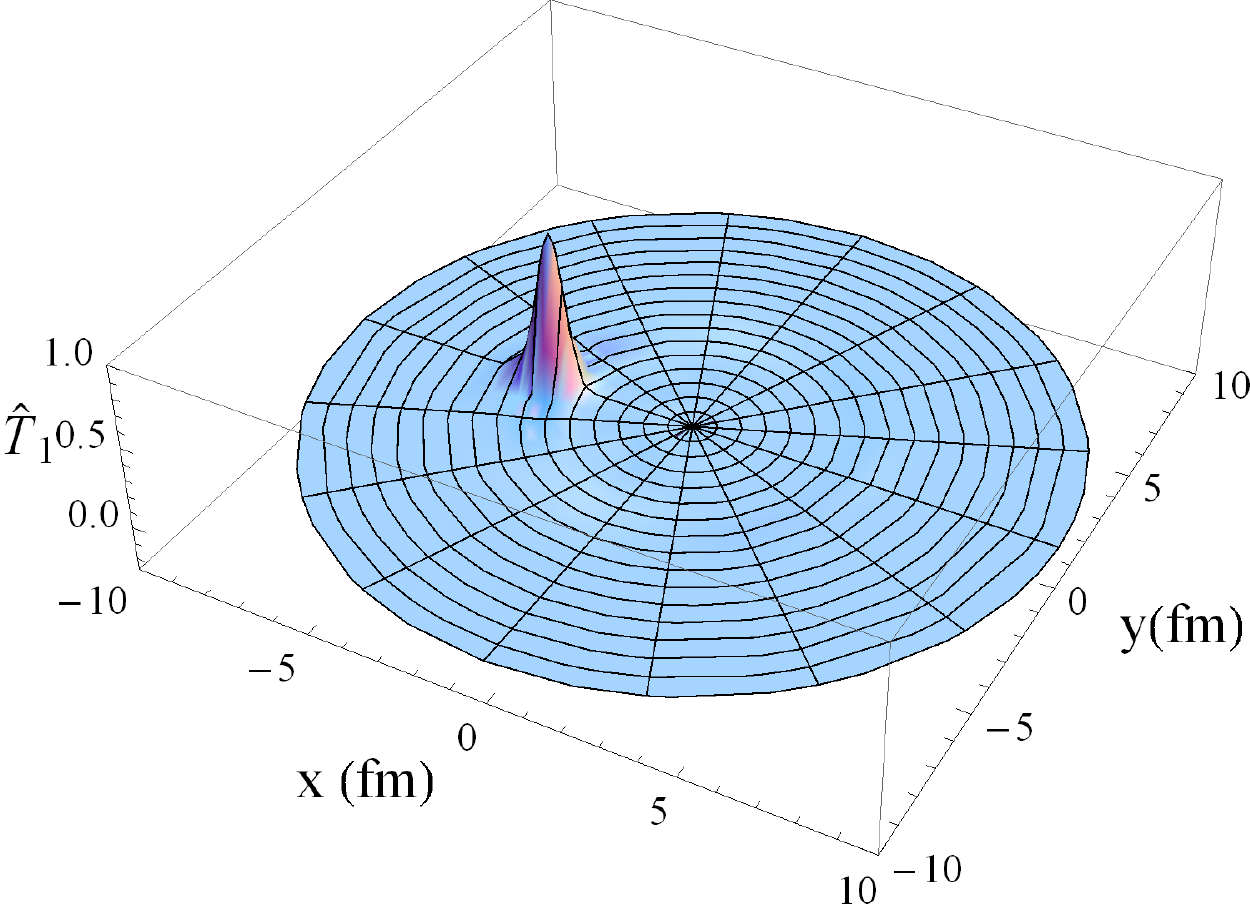}
\includegraphics[width=8 cm]{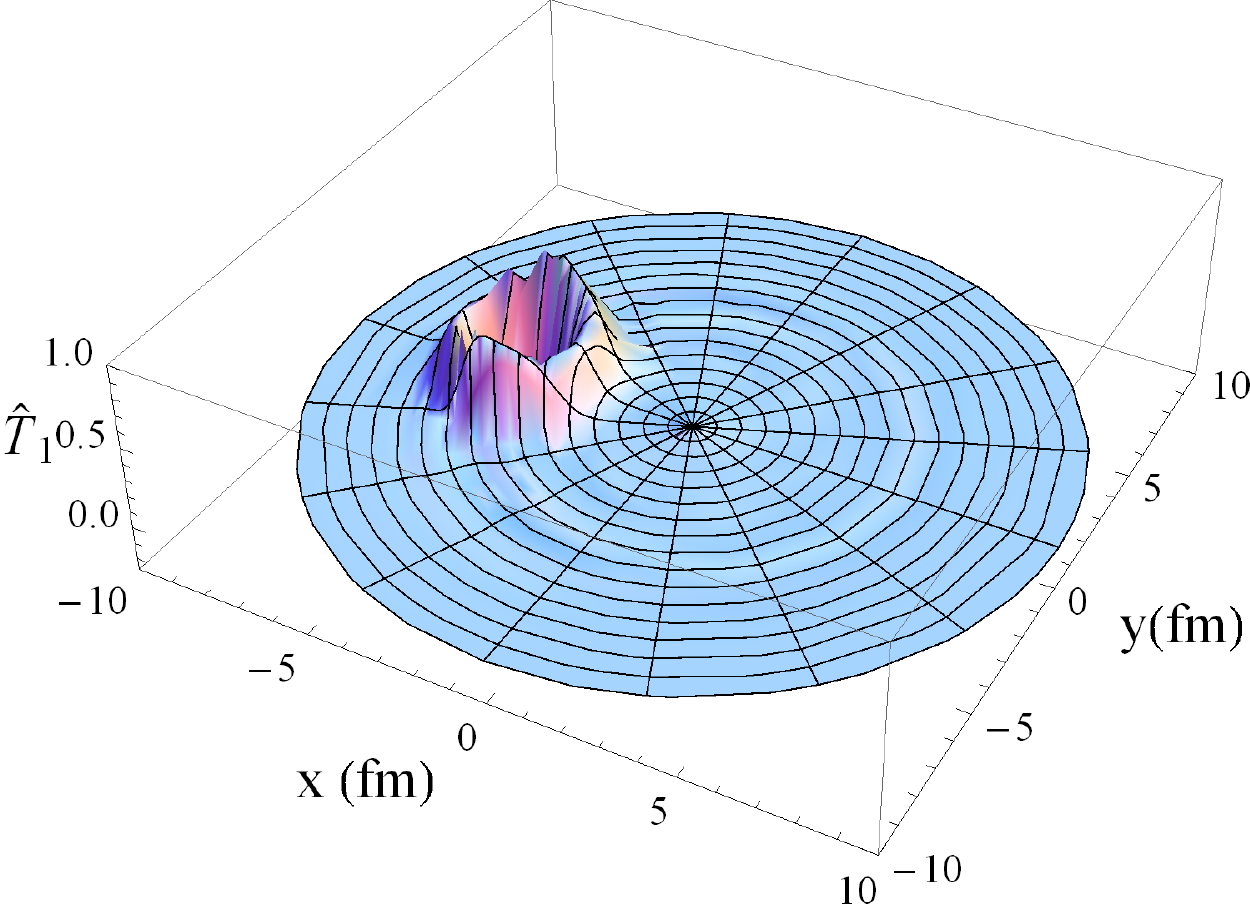}
\includegraphics[width=8 cm]{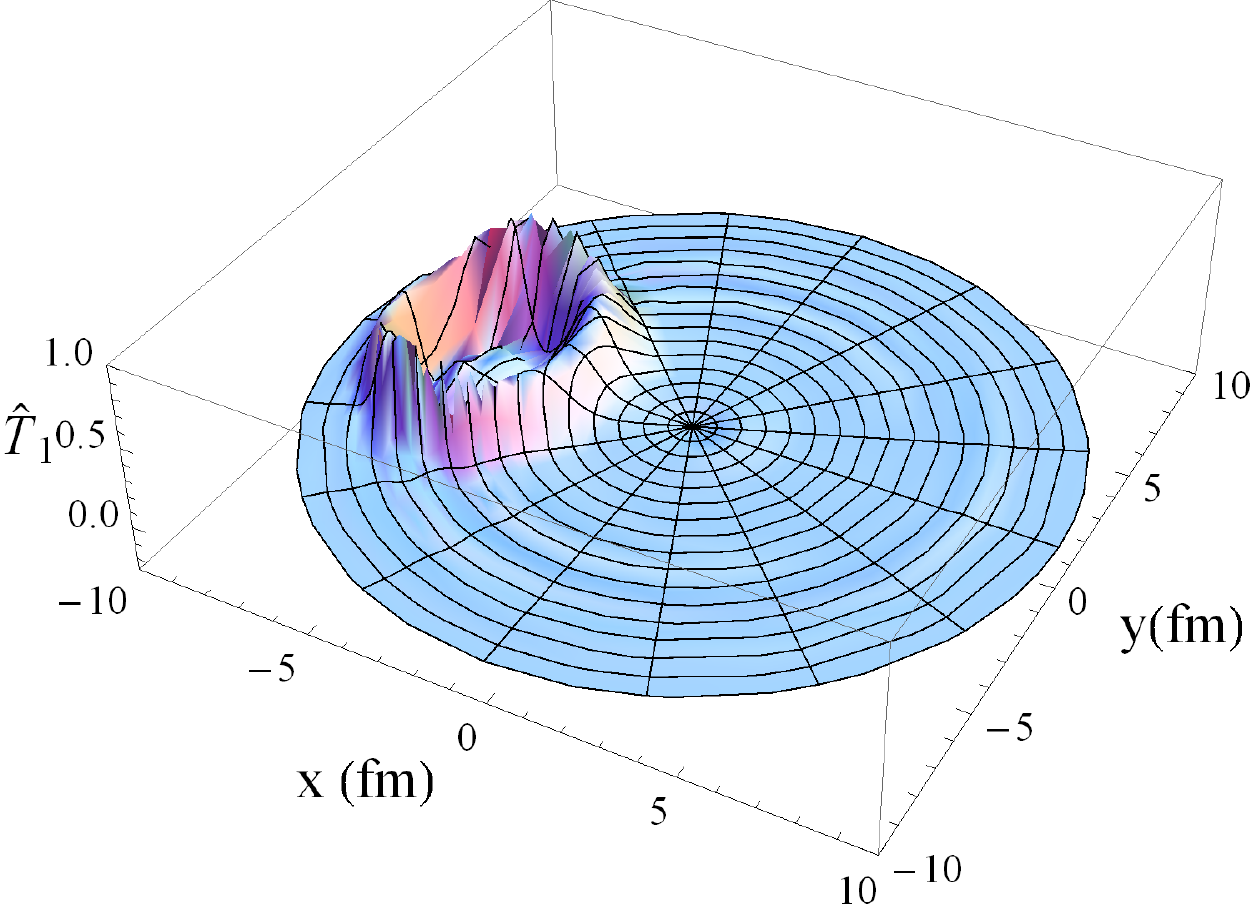}
\end{center}
\vspace{-5ex}\caption{(Color online) Evolution of the perturbation in the rescaled frame but in the regular coordinates $\hat{T}_1(\tau,r,\phi)$
from (\ref{Tpert}) using the change of coordinates
\ref{rho_coord} and \ref{theta_coord}.
 From top to bottom:$\tau=1\, fm/c$ ,$\tau=4\, fm/c$ ,$\tau=6 \, fm/c$}\label{t_evol}
\end{figure}

The $c_{lm}$ coefficients are calculated using the orthogonality
of the Legendre polynomials, and are given by:
\begin{eqnarray}
c_{lm} & = & \int_0^{2\pi}\int_0^{\pi}
\hat{T}_1(\rho_0,\theta,\phi)
Y_{lm}^{*}(\theta,\phi)\sin{\theta}d\theta d\phi \nonumber\\
\end{eqnarray}
Once we find all the constants, we can study the evolution of the perturbation given by expression (\ref{Tpert}). In figure
\ref{t_evol} we show three frames from a movie-like  evolution in $\tau$ of a perturbation $\hat{T}_1(\tau,r,\phi)$ produced by a local ``hot spot" which was calculated for a Gaussian centered in $\theta=1.5$  with a small size $s=0.1$, which corresponds to a perturbation localized at $r=4.1$ fm and with a width of 0.4 fm. Notice that while the perturbation is in the rescaled frame, we are using the regular coordinates $\tau, \, r $.

We have used 30 harmonics for this movie, and it is nice to see that they all add up coherently into a consistent picture of a sound wave propagation. While it does correspond to a qualitative picture of a circle from a stone thrown into the pond, with which we had started this work, it is in fact an exact solution, riding on the zeroth order explosion picture which is by itself rather complex.
In order to find the analytical expressions for the perturbation
on top of the fireball it was necessary to invent the $\rho$ and
$\theta$ coordinates, so that all of the expressions can be factorized in terms of these coordinates. So a lot of correct thinking was needed, to make this movie possible.

\subsection{The viscous effects}
In the second  paper of this series \cite{Staig:2010pn} we introduced the viscosity-based scale, which all structures produced by point-like perturbations would obtain at freezeout. Without going into details, let us just remind the reader that while  the width of the circle grows with time as $\tau^{1/2}$, its radius  grows as $\tau$, and therefore the relative contrast (the former divided by the latter) is improving as $\tau^{-1/2}$. As far as the amplitude of the wave is concerned,  in a short-plain-wavelength approximation the stress tensor harmonics with momentum $k$ are attenuated by a factor
\be %
\delta T_{\mu\nu} (t,k) = exp\left(-{2 \over 3} {\eta \over s}
{k^2 t \over  T } \right)  \delta T_{\mu\nu} (0,k)
\label{eqn_visc_filter} \ee %
known from textbooks on sound, sometimes called  ``the viscous
filter". Note that its exponent contains the momentum $squared$,
due to the extra derivative in the viscous tensor, and therefore
the effect of viscosity  for the higher harmonics is strongly
enhanced. Obviously, the same qualitative behavior is expected for our $l,m$ harmonics.

The basic equations for the $\rho$-dependent part of the perturbation, now with viscosity terms, can be written as a system of coupled first-order equations \cite{Gubser:2010ui}. We are  assuming rapidity independence, thus the system of equations $(107)$,$(108)$ and $(109)$, from the referred paper, becomes two coupled equations, for  (the $\rho$-dependent part of) the temperature and velocity perturbations
\begin{eqnarray}
\frac{d\vec{w}}{d\rho}= -\Gamma \vec{w} \,\, ,\,\,\,\,\,\, \vec{w} & = &\left(%
\begin{array}{c}
  \delta_v \\
  v_v \\
\end{array}%
\right) \label{visMeq}
\end{eqnarray}
where the index $v$ stands for viscous and the matrix components
are, \small
\begin{eqnarray}
\Gamma_{11} & = & \frac{H_0 \tanh^2{\rho}}{3\hat{T}_b} \nonumber \\
\Gamma_{12} & = & \frac{l(l+1)}{3\hat{T}_b \cosh^2{\rho}}
\left(H_0\tanh{\rho}-\hat{T}_b\right)\label{visMatrix}\\
\Gamma_{21} & = & \frac{2H_0 \tanh{\rho}}{H_0\tanh{\rho}-2\hat{T}_b}+1 \nonumber \\
\Gamma_{22} & = & \frac{ 8\hat{T}_b^2\tanh{\rho} + H_0\hat{T}_b
\left( \frac{-4( 3l( l+1 )- 10 )}{ \cosh^2{\rho}}-16 \right) +
6H_0^2\tanh^3{\rho} }{ 6\hat{T}_b \left( H_0\tanh{\rho} -
2\hat{T}_b \right) }\nonumber
\end{eqnarray}
\normalsize %
This system can also be written as a closed second order differential equation for $\delta_v(\rho)$: \small
\begin{eqnarray} \label{viseqn}
\frac{d^2\delta_v}{d\rho^2}-\frac{d\delta_v}{d\rho}\left(\Gamma_{11}-
\frac{1}{\Gamma_{12}}\frac{d\Gamma_{12}}{d\rho}+\Gamma_{22}\right) \\
-
\delta_v\left(\frac{d\Gamma_{11}}{d\rho}-\frac{\Gamma_{11}}{\Gamma{12}}\frac{d\Gamma_{12}}{d\rho}-\Gamma_{11}\Gamma_{22}+\Gamma_{12}\Gamma_{21}\right)
& = & 0\nonumber
\end{eqnarray}

\normalsize \noindent
Unfortunately, unlike the zero viscosity case considered above, the equations one gets after separation of variables cannot all be solved analytically and thus have to be solved numerically which has been done using Mathematica's ODE solver. The part of the solution which depends on $\theta$ and $\phi$ is not affected by viscosity, so it continues to be given by the spherical harmonics $Y_{lm}(\theta,\phi)$.

Our results for the nonzero viscosity will use either $H_0=0.33$ ($\eta/s=0.134$), like in \cite{Gubser:2010ui},  or the value $H_0=0.19$ ($\eta/s=1/(4\pi)=0.08$), the conjectured lowest value possible predicted by AdS/CFT in the strong coupling limit.

\begin{figure}[!h]
\begin{center}
\includegraphics[width=6 cm]{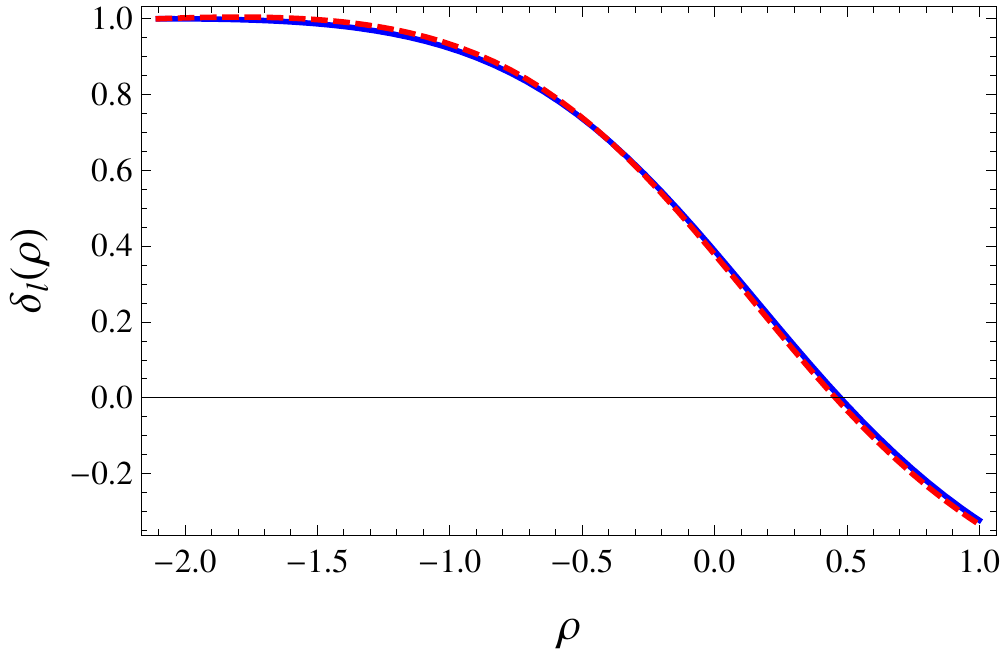}
\includegraphics[width=6 cm]{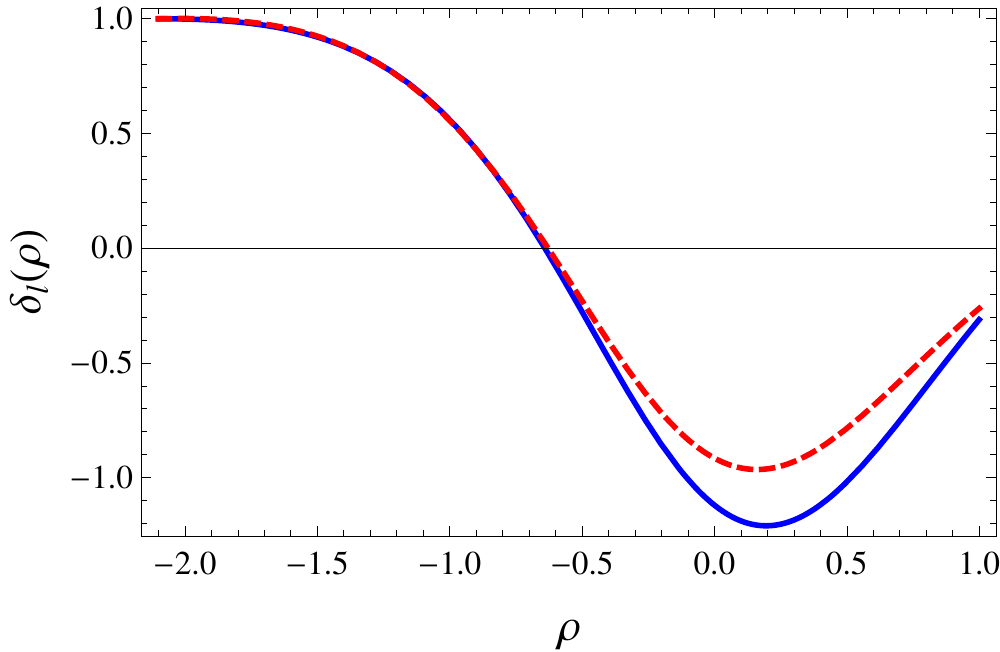}
\includegraphics[width=6 cm]{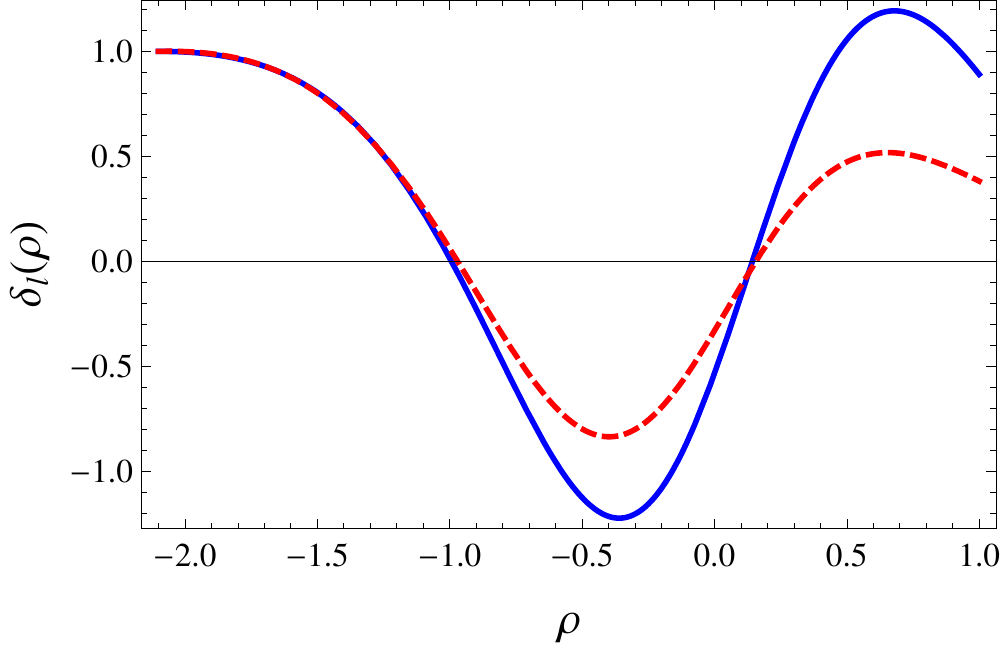}
\includegraphics[width=6 cm]{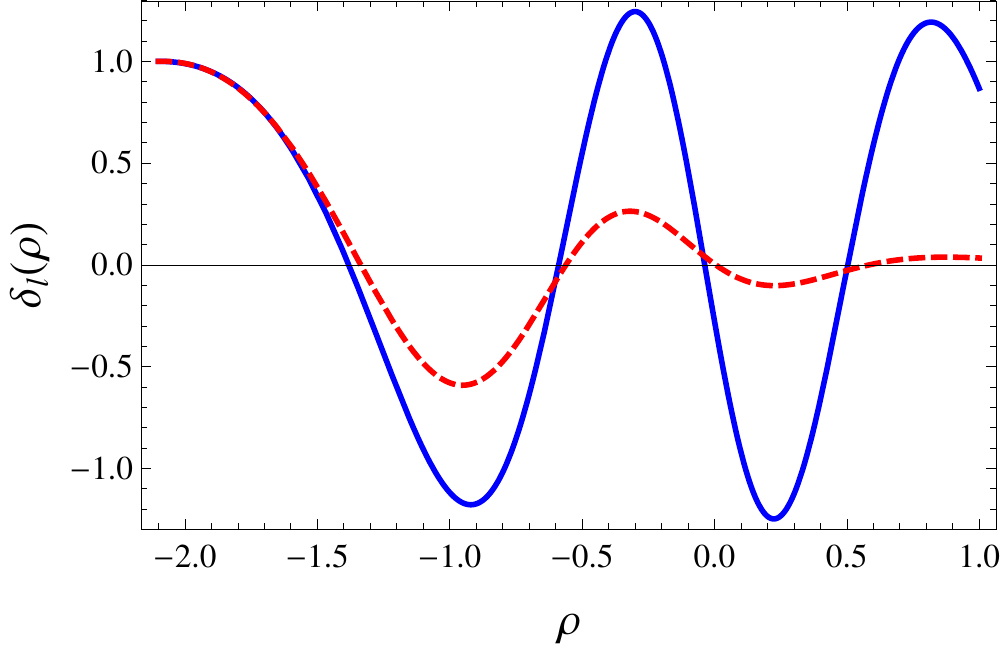}
\end{center}
\vspace{-5ex}\caption{(Color online) Comparison between the magnitude of harmonics  $\delta_l (\rho)$
(\ref{deltaI}) for the ideal case (solid blue lines) and $\delta_{v \,
l}(\rho)$ (\ref{viseqn}) for the viscous case with
$\eta/s=0.134$ (dashed red lines), for $l=1,3,5,10$, from top to
bottom.}\label{delta_com}
\end{figure}

 In Fig.\ref{delta_com} we plot the ``time" dependence $\delta_{v \, l}(\rho)$ for several harmonics and compare  them to the inviscid case $\delta_{l}(\rho)$ for some l's. As expected, the viscosity reduces higher harmonics more, but as far as time dependence is compared to inviscid case, we see that viscosity literally kills the contribution at certain time, which becomes shorter and shorter for  larger $l$.  As the time is limited by the freezeout time, we observe that  the contributions  of all sufficiently large $l > l_{max}\sim 10$ become completely negligible.

\begin{figure}[!h]
\begin{center}
\includegraphics[width=8 cm]{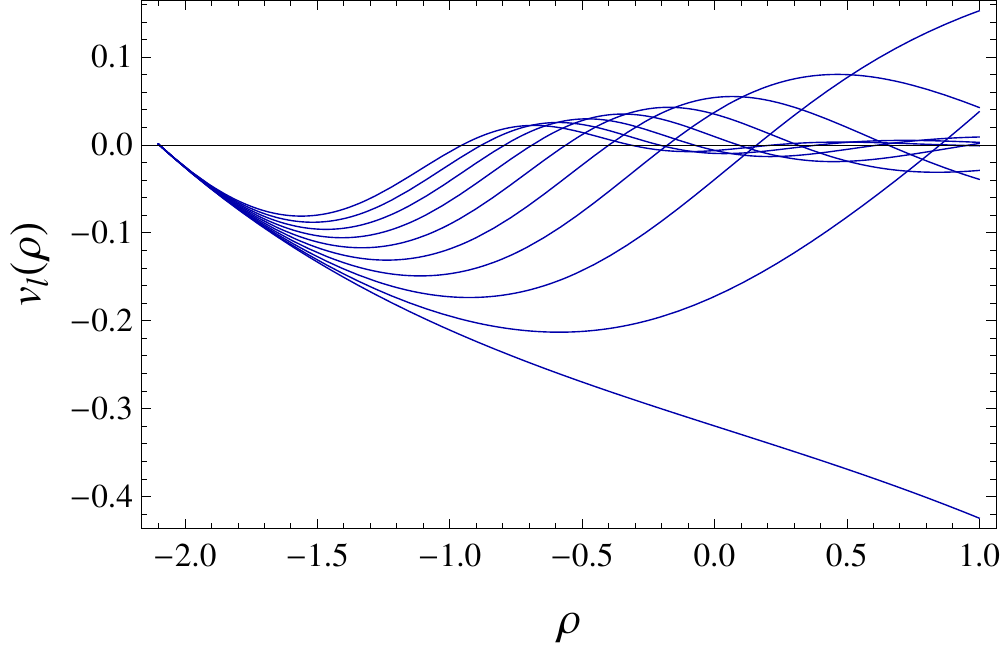}
\end{center}
\vspace{-5ex}\caption{(Color online) The velocity harmonics $v_{v \,l}(\rho)$ (defined in (\ref{vv})) for the first ten values of  $l$, in the viscous case with
$\eta/s=0.134$. To be compared with figure \ref{amplitudes}
bottom.}\label{vlvis}
\end{figure}

The $\rho$-dependent part of the velocity can be calculated once
$\delta_v$ is known:
\begin{eqnarray}
v_{v \, l} (\rho) & = & -\frac{\delta'_{v \,
l}+\Gamma_{11}\delta_{v \, l}}{\Gamma_{12}}\label{vv}
\end{eqnarray}
For the first $10$ l, the curves $v_{v \, l} (\rho)$ are plotted
in Fig.\ref{vlvis}. Comparing this to $v_l(\rho)$ at zero viscosity (bottom plot of Fig.\ref{amplitudes})  we see that the amplitude for the velocity is also damped in the viscous case for large l and increasing
$\rho$.
\section{Applications}

\subsection{Matching the Gubser flow with the heavy ion collisions}

With the exact solution to the perturbation equation riding on top of the Gubser flow at hand, one may  go back to the $\tau$
and $r$ coordinates  and try to calculate what should happen in
real heavy ion collisions.

But before we do so, let us remind the reader once again that the Gubser flow is by itself an idealization of reality.  The real hadronic matter  can only be approximated by the conformal EOS $\epsilon=3p$ during its QGP phase, which lasts about 1/3 of the total time at RHIC and perhaps around 1/2 time at LHC. The rest is the near-$T_c$ domain and the hadronic phase, in which the speed of sound changes from $1/\sqrt{3}=.577$ to about $.35$ and $.45$, respectively. Although this change is not very large, we do notice that the radial  flow obtained with the Gubser flow is too large. Respectively, the freezeout time $\tau_{FO}$ is indeed somewhat smaller than that observed in numerical hydrodynamics with correct EOS. Perhaps some of our results for the perturbation should also need some adjustment, due to these facts.

The second similar comment is that Gubser's solution has a particular shape, which has no reason to coincide with the shape of the real Au nuclei.  The finite size of the fireball and the shape of its temperature profile is determined by the parameter q which we take equal to $(4.3 fm)^{-1}$ following \cite{Gubser:2010ze}.  The second parameter that we need to fix is the constant $\hat{T}_0$.  Again from \cite{Gubser:2010ze} we get the formula for this parameter
\begin{eqnarray}
\hat{T}_0 = \frac{1}{f_*^{1/12}}\left(\frac{3}{16 \pi}\frac{dS}{d\eta}\right)
\end{eqnarray}
with
\begin{eqnarray}
f_* = \frac{\epsilon}{T^4} = 11, \;\;\;\; \frac{dS}{d\eta} =7.5 \frac{dN_{ch}}{d\eta}
\end{eqnarray}
For central $(0-5\%)$ collisions at LHC $dN_{ch}/d\eta \sim 1600$ \cite{Aamodt:2010cz} which gives a value of $\hat{T}_0 \approx 7.3 $.  Using these values one gets a freeze-out time $\tau_{fo} \sim 6$, which is rather a short time that doesn't allow for the sufficient evolution of the sound circles.  Since we are interested in studying the propagation of sound perturbations and the size of the sound horizon depends on the freeze-out time, we will use $\hat{T}_0 \approx 10.1$, which corresponds to having about $2.6 (dN_{ch}/d\eta)_{LHC}$. It is important to note that the background temperature in the ideal case corresponds to
\begin{eqnarray}
T=\frac{1}{\tau f_*}\frac{\hat{T}_0}{(\cosh{\rho})^{2/3}},
\end{eqnarray}
so we are using an initial temperature of about 630 MeV. The parameters we used are such that the size of the fireball at freeze-out, the radius of the sound circle and overall transverse expansion velocities $v_\perp(r, t\approx t_f) $ mimic reality of RHIC/LHC collisions. The price for that is somewhat too large initial temperature and overall entropy. 

The hydrodynamical equations should be used only after some
approximate equilibration of hadronic matter is achieved. While
the mechanism of it, as well as precise timing remains unknown, we do know its order of magnitude to be a fraction of fm/c. For our calculations we assume that thermalization occurs at the initial time $\tau=1 \, fm/c$, and it is at this time that we define our initial ``hot spot" and start evolving it using 
hydrodynamics. One can do so until the final freeze-out is
reached, at which point the interaction between secondaries becomes ineffective and sound propagation stops. Below we will discuss how the hydrodynamical perturbations should be translated into the experimental observables.

Let us point out that we study the effect of a single hot-spot on the fireball which we characterize as a Gaussian temperature
perturbation on top of the background temperature. In real
collisions, there are many such perturbations, but since we solve the problem in the linear approximation, their evolution is mutually independent. Furthermore, in the experimental statistical study of small two- or three-particle correlations, the contribution of the uncorrelated fluctuations is cancelled out automatically.

In Fig.\ref{t_evol} we see that at the time $\tau=1 \, fm/c$  a
Gaussian ``hot spot" centered at $r=4.13 fm, \, \phi=\pi $
corresponds to having it at ``time" $\rho=-2.07$  centered at
$\theta=1.5, \, \phi=\pi $. Of course, since $\rho=\rho(\tau,r)$
at any given time $\tau$, $\rho$  depends on $r$, so the initial
condition $\rho=-2.07$ is for the center of the Gaussian.

\subsection{Modification of the freezeout surface and of the particle spectra}

The standard expression for a spectrum, known as Cooper-Frye formula \cite{Cooper:1974mv}, is given by%
\be  E{dN\over d^3p}= - \int d\Sigma_\mu p^\mu
f\left( {p^\nu u_\nu \over T}\right). \ee%
The overall minus is there because we work using the mostly plus
signature. The function f corresponds to the thermal distribution inside the fluid cells, boosted by their hydrodynamical motion at the time of the freezeout%
\be f(p)= {1 \over exp(-p^\mu u_\mu/T)\pm
1} \ee%
for Bose/Fermi particles. (In reality, we will be only interested in the tail, so the Boltzmann approximation will always be enough.) The minus sign in the exponent is because we are working in the mostly plus signature.

The temperature and velocity in this formula are supposed to have a space-time dependence derived from hydrodynamics. The freeze-out surface $\Sigma^{\mu}$ that appears in the Cooper-Frye formula corresponds to a certain kinetic condition, of the form that the ratio of a particular reaction rate to the matter expansion rate reaches a particular value. Since there are many reactions involved in the process, strictly speaking there are multiple freezeout surfaces. One usually separates ``chemical" and ``kinetic" freezeouts, in which inelastic and elastic scattering rates are involved. Since different secondaries (pions, K, nucleons, ... $J/\psi$) in fact have quite different elastic cross sections, the ``kinetic" surfaces should in fact be different for each species.

We are not going to discuss all those complications in this work, and think of only one type of secondaries, the pions. Furthermore, we will use a drastic  simplification often used, assuming that the freezeout surface is the $isotherm$ $T(t,x)=T_{FO}$. If so, the surface can be determined from hydrodynamical output, for example its time-like part can be written as
\begin{eqnarray}
\Sigma^{\mu} & = & (\tau_{fo}(x,y),x,y,\eta)
\end{eqnarray}
where $\tau_{fo}$ is the time at which the fireball reaches the
freeze-out temperature. The Cooper-Fry formula contains the vector normal to the surface which is then
\begin{eqnarray}
d\Sigma_{\mu} & = & -\sqrt{-g}\epsilon_{\mu \nu \lambda
\rho}\frac{\partial \Sigma^{\nu}}{\partial x} \frac{\partial
\Sigma^{\lambda}}{\partial y} \frac{\partial
\Sigma^{\rho}}{\partial \eta} dxdyd\eta \\
& = & \left(-1,\frac{\partial \tau_{fo}}{\partial
x},\frac{\partial \tau_{fo}}{\partial y},0\right)\tau_{fo}
dxdyd\eta
\end{eqnarray}
Here $g$ is the determinant of the metric and $\epsilon_{\mu \nu
\lambda \rho}$ is the Levi-Civita symbol.

The perturbations affect the spectra in two ways. First, the flow  velocity in the exponent is corrected by the extra terms of the first order due to sound. The second effect, related with the first order  temperature perturbations $(1+\delta)$, are more subtle. Hotter matter (positive $\delta$) in the event with a ``hot spot" and perturbation from it imply a production of extra entropy density (increases by $(1+\delta)^3$)  as compared to the zeroth order fireball. This means there would be extra secondaries produced, as this entropy is ``hadronized".   By assumption, it happens locally, delaying a bit the freezeout according to condition \be T_0(t,x)\left[ 1+\delta(x,t) \right]=T_{FO} \ee Thus delay is absolutely necessary, it provides  extra volume for the extra  matter produced, as compared to the zeroth order explosion, since by assumption the freezeout  temperature and thus the matter density at the FO surface are held constant. The deformation of the FO surface not only increases the volume, giving place for the extra particles just discussed, but it also prolongs hydro evolution, providing a bit larger flow.

Let us now discuss another issue: at what part of the particle
spectra we should focus, in order to see best the effect of the
perturbation. The Cooper-Fry  formula has $p_t$ of the particle in the exponent, so it is tempting to take it as large as possible. And indeed, all hydro effects (such as e.g. the elliptic or radial flow) are  enhanced by the increase in the particle momentum $p_t$. There are two practical limits to  an increase in $p_t$, however:\\
(i) One can be understood inside the hydrodynamics itself. The
viscous term has an extra gradient, relative to the ideal part of the stress tensor. This means that the relative role of  viscous corrections will grow with $p_t$, till at some point it will no longer be small as compared to ideal term. Obviously at such $p_t$ hydrodynamics should be substituted by some other tool, e.g. some kinetic theory description.\\
(ii) In real collisions  some secondaries originate from hard
scattering and subsequent jets. In spite of significant jet
quenching, at large enough $p_t$ the hard component of the spectra supersedes the hydrodynamical spectra. Obviously, beyond this point one looses ability to follow the hydrodynamical component.
 
The transition between the hydrodynamic part of the spectrum and the hard QCD tail  has been determined to be  between 4-5 GeV \cite{Fries:2003vb,Renk:2011gj} so, a bit conservatively, we will consider  $p_t=1 \, GeV$, as a region well inside the hydrodynamical domain. Even at this $p_t$, its ratio to the kinetic FO temperature is a large number $p_t/T_f=O(10)$, which  can be treated as a large parameter of the problem, residing in the exponent.

Let us work out the first-order corrections appearing from the
perturbation. There are two effects, one from the extra matter
$T=T_f+\delta T$ and one from extra motion of the matter in the
sound wave. The latter contribution comes simply from adding the
perturbation to the velocity,
\begin{eqnarray}
u_{\mu}\rightarrow u_{\mu}+\delta u_{\mu}
\end{eqnarray}
$\delta u_{\mu}$ is the perturbation, written in (\ref{vpert}) as
$\hat{u}_1$ times $\tau$.

\begin{figure}[!h]
\begin{center}
\includegraphics[width=8 cm]{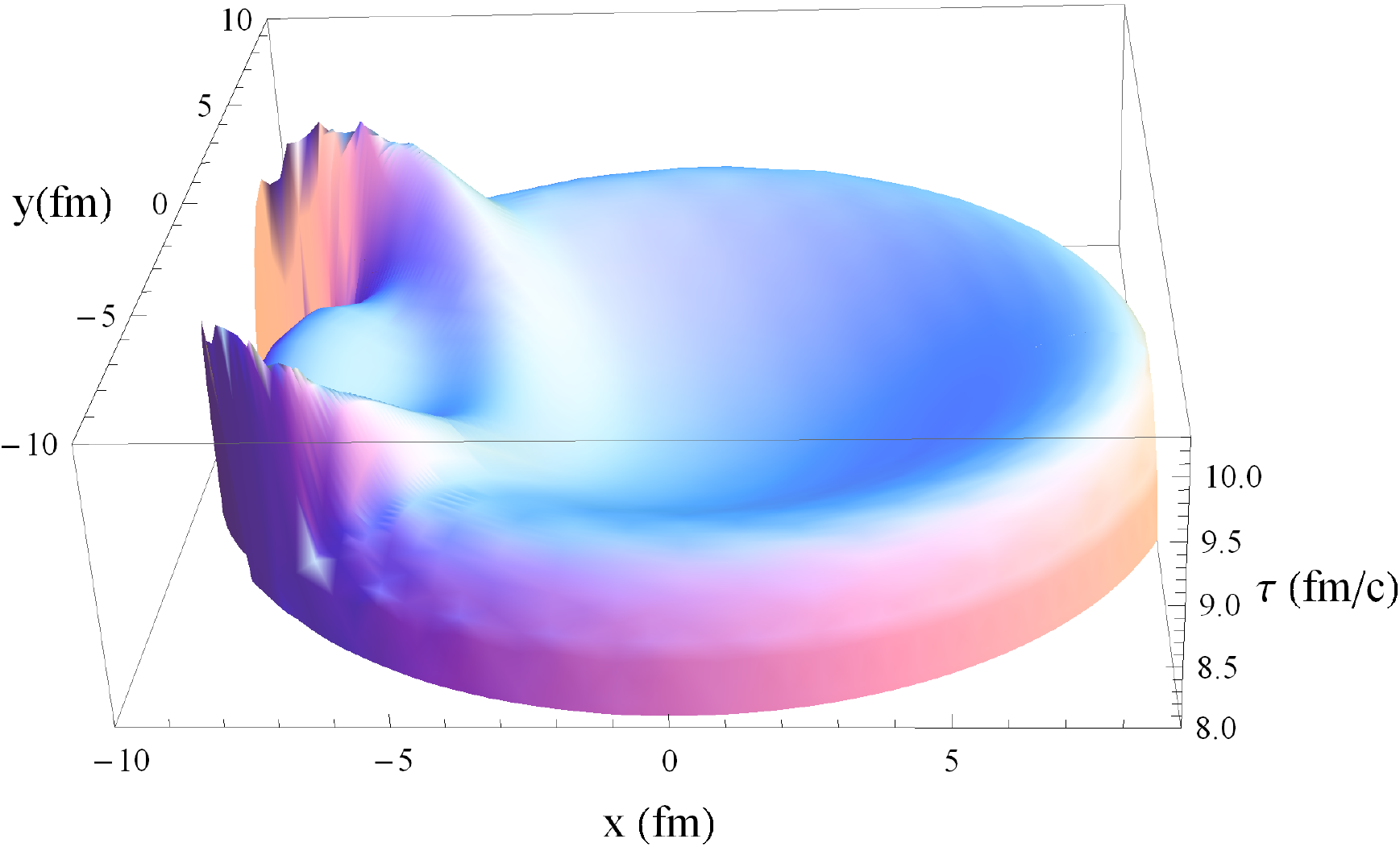}
\end{center}
\vspace{-5ex}\caption{(Color online) Freeze-out surface $\tau(x,y)$ for the
inviscid case.}\label{foS_invis}
\end{figure}

\begin{figure}[!h]
\begin{center}
\includegraphics[width=8 cm]{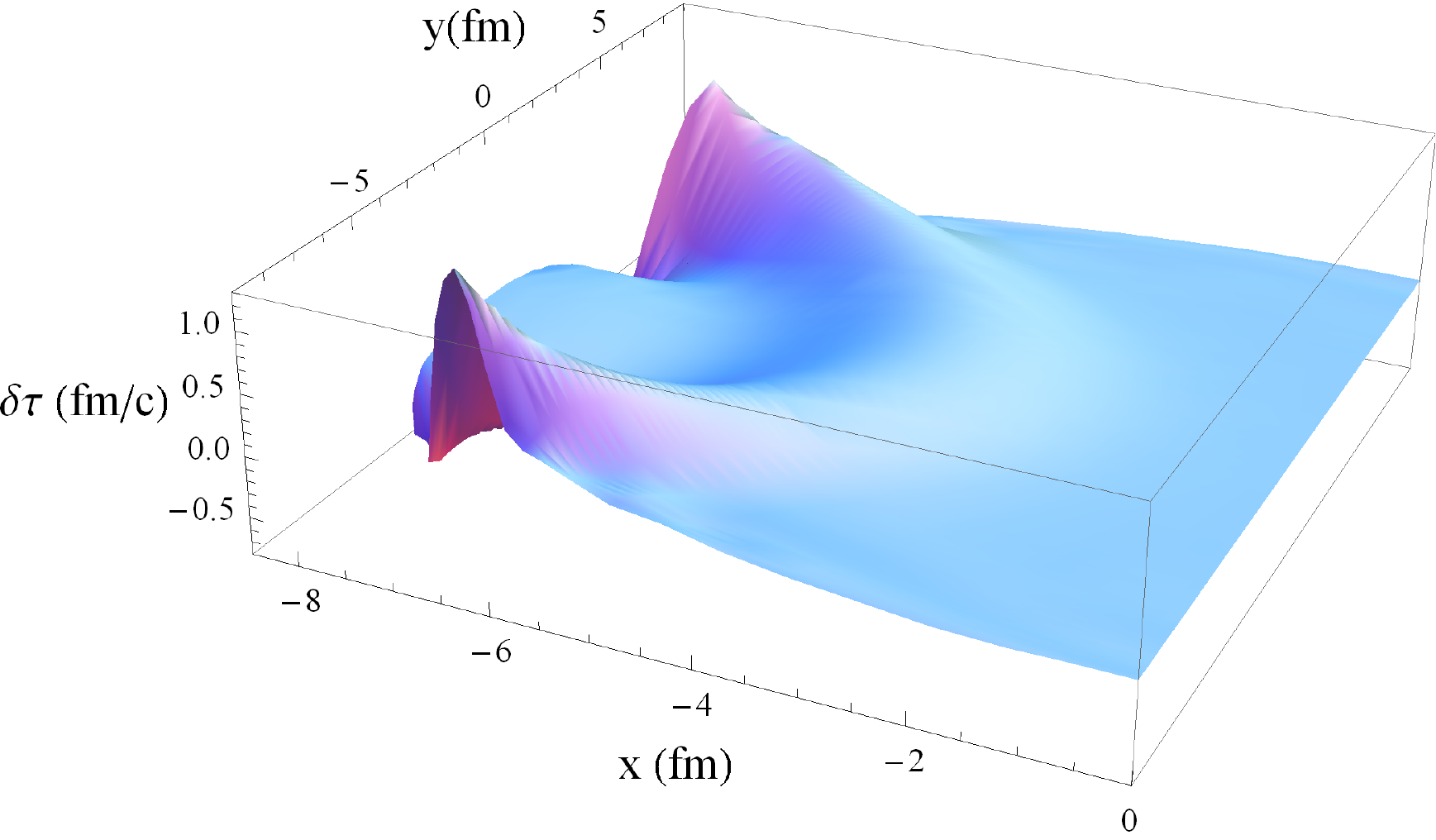}
\includegraphics[width=8 cm]{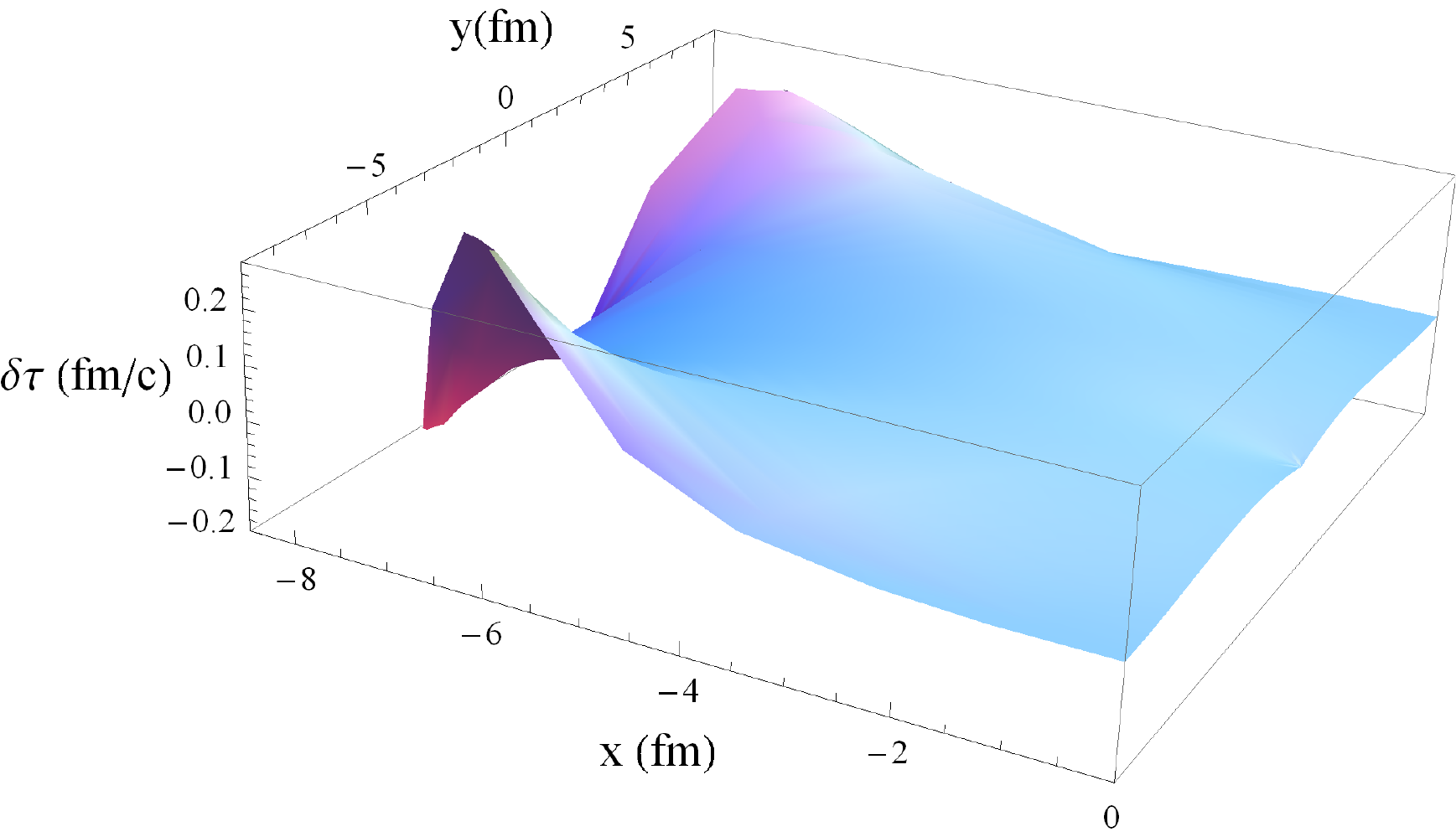}
\end{center}
\vspace{-5ex}\caption{(Color online) Excess of freeze-out surface
$\delta \tau(r,\phi)$ due to the initial perturbation. Top: ideal
case, bottom: viscous case with $\eta/s=0.134$. Only the half of
the surface that is affected by the presence of the perturbation
was plotted.}\label{dtau}
\end{figure}

The effect due to the extra matter is included when calculating
the freeze-out surface:
\begin{eqnarray}
T_{fo} & = & T_b(\tau,r) + \delta T(\tau,r,\phi)\label{tfo_eqn}
\end{eqnarray}
where $\delta T=\hat{T}_1/\tau$, with $\hat{T}_1$ from
(\ref{Tpert}).The equation (\ref{tfo_eqn})  is solved for
$\tau(r,\phi)$, and the result for the inviscid case is presented
in Fig.\ref{foS_invis}.
Since the contribution from the perturbation is small, we write
$\tau(r,\phi)=\tau_b(r)+\delta\tau(r,\phi)$ and consider terms up to first order in $\delta\tau(r,\phi)$. By this we mean that the exponent will be approximated by \small
\begin{eqnarray}
\frac{p^{\mu}u_{\mu}(\tau_b+\delta\tau)}{T_f} & \approx &
\frac{p^{\mu}u_{b \,\mu}(\tau_b)}{T_f}
+\frac{1}{T_f}\frac{d(p^{\mu}u_{b \,
\mu}(\tau_b+\delta\tau))}{d(\delta\tau)}|_{\delta\tau=0}\delta\tau \nonumber\\
{} & & +\frac{p^{\mu}\delta u_{\mu}(\tau_b)}{T_f}
\end{eqnarray}\normalsize
Fig.\ref{dtau} shows $\delta\tau$ for both, the inviscid and for
the viscous case. In the former case the contribution is much
larger than in the latter, where the viscosity has damped and
widened the peaks.

\begin{figure}[!t]
\begin{center}
\includegraphics[width=7 cm]{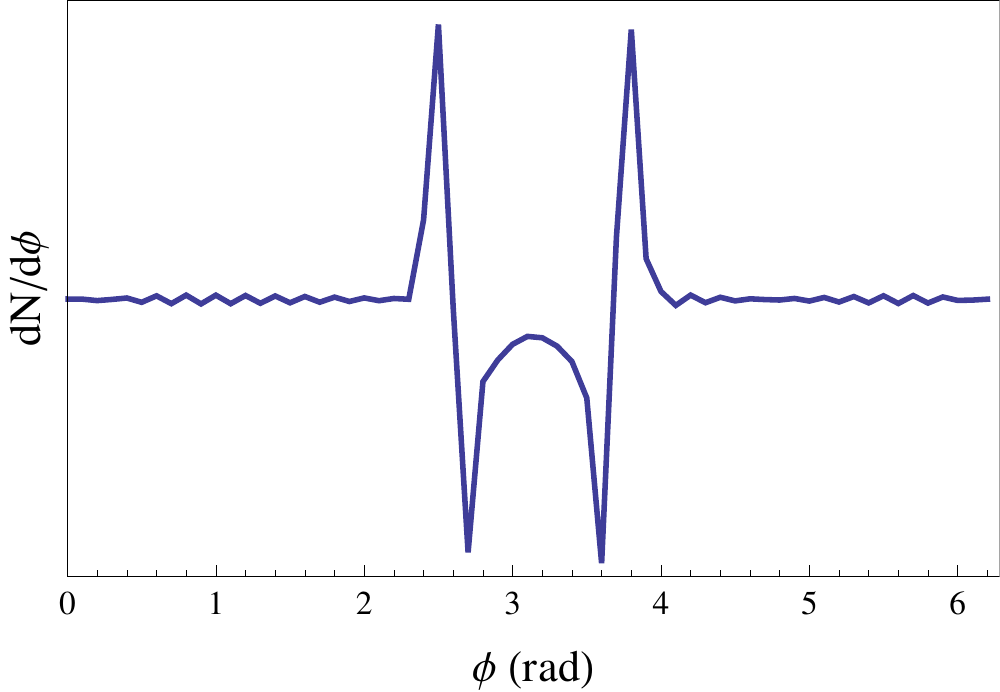}
\includegraphics[width=7 cm]{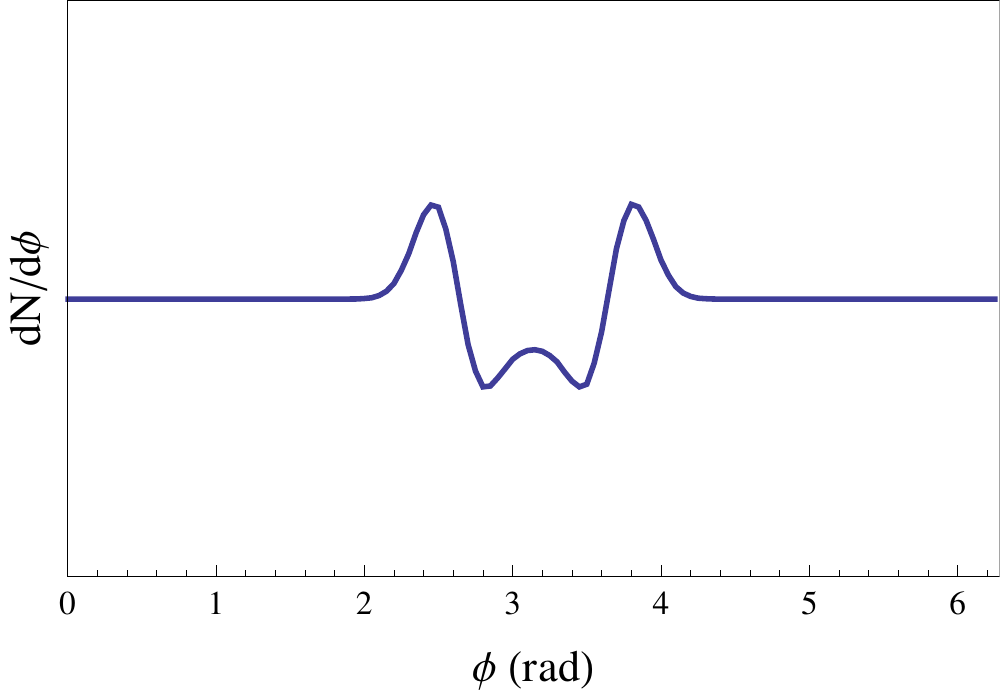}
\includegraphics[width=7 cm]{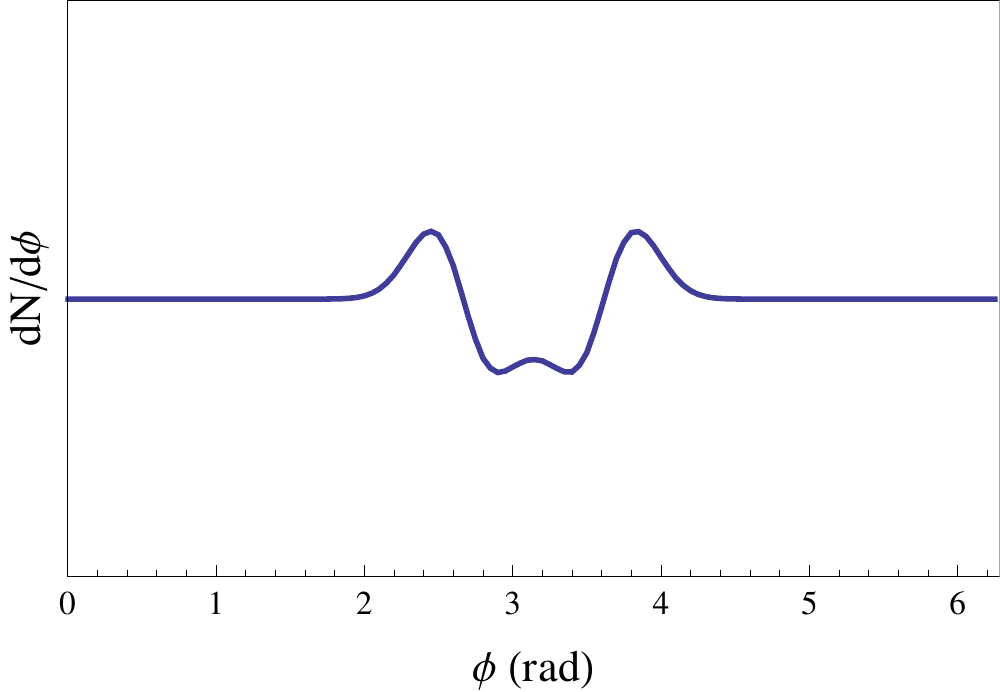}
\end{center}
\vspace{-5ex}\caption{(Color online) Single-pion distribution in arbitrary units as a function of the azimuthal angle $\phi$ (rad),
for transverse momentum $p_T=1\,GeV$ and rapidity $y=0$. From top to bottom, the curves are for different viscosity-to-entropy ratios,  $\eta/s = 0,\,
1/(4\pi), \, 0.134$ respectively.}\label{pdist}
\end{figure}
Figure \ref{pdist} compares the particle distribution for 
three cases, (i) the inviscid case, (ii) the minimal  viscosity case $\eta/s=1/(4\pi)$ and (iii) the case where
$\eta/s = 0.134$.
In the ideal hydro case the two peaks of the angular
distributions, due to the overlap of the perturbation with the
fireball boundary, are more pronounced than in the cases with
nonzero viscosity. Also, in this case (i) one can clearly see high frequency oscillations  on the curve.  Those are an artifact of the arbitrary limit of the number of harmonics used to $l<l_{max}=30$. The oscillations disappear when we take viscosity into account, because, as we mentioned earlier, viscosity kills all higher harmonics anyway, with $l>\l_{max}\sim 10$. In the presence of viscosity, the peaks in the particle distribution are weakened, and their angular separation is a bit more spread than in  the inviscid case.

\subsection{Two-particle correlations }

\begin{figure}[!h]
\begin{center}
\includegraphics[width=6 cm]{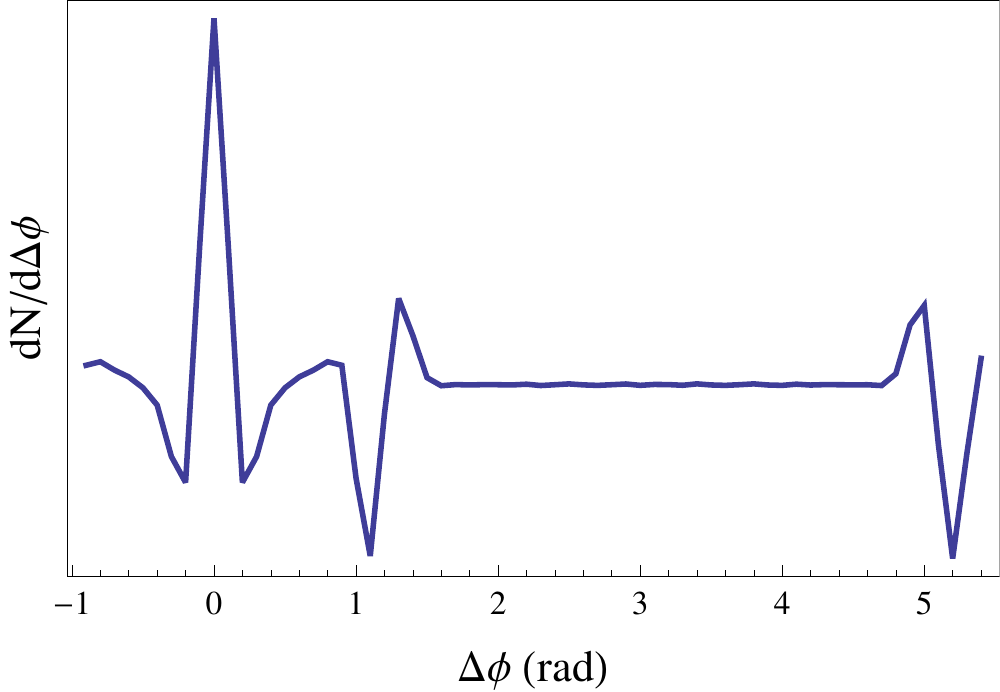}
\includegraphics[width=6 cm]{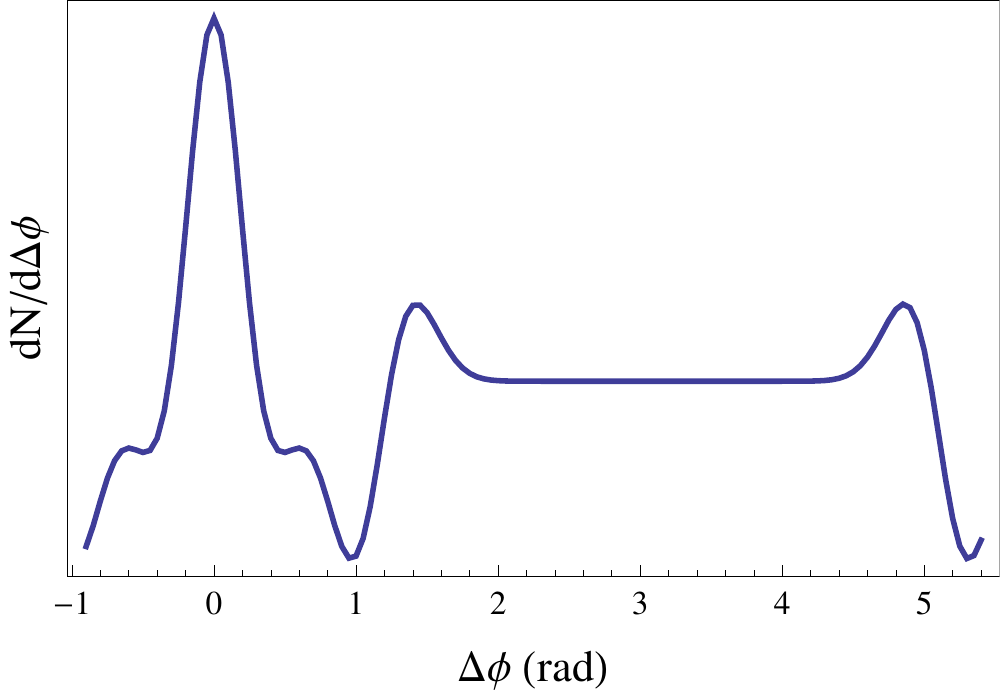}
\includegraphics[width=6 cm]{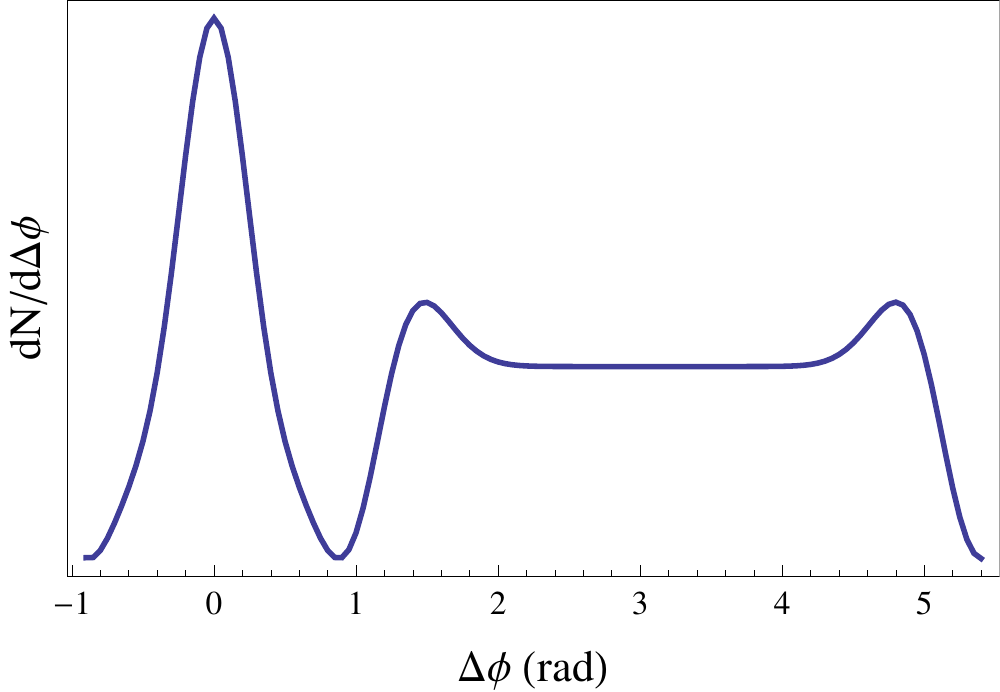}
\includegraphics[width=6. cm]{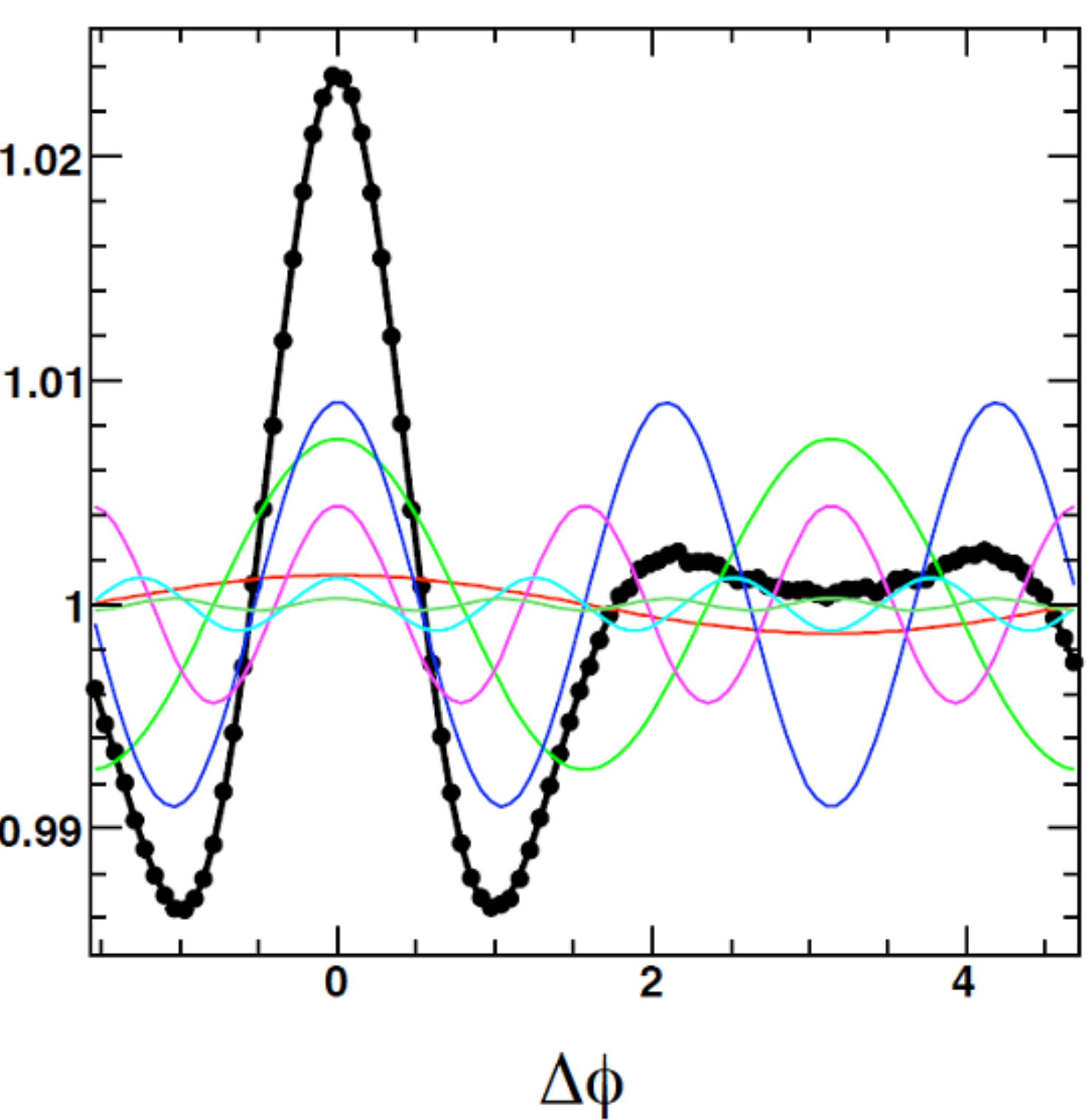}
\end{center}
\vspace{-5ex}\caption{(Color online) The two-pion distribution in arbitrary units as a function of azimuthal angle difference $\Delta\phi$ (rad). From top to bottom, the three upper plots correspond to calculated shapes, for values of the
viscosity-to-entropy ratios, $\eta/s=0,\,1/4\pi,\,0.134$,
respectively.
The bottom plot, shown for comparison, is measured at LHC by ATLAS collaboration \cite{ATLAS_corr}, for the most central (0-1\%) collisions. Very similar shape of the correlation function is in fact observed by all five collaborations at RHIC and LHC.
}
\label{2pdist}
\end{figure}

Looking at experimental data on normalized two particle correlations, such as the one shown in the last plot of Fig.\ref{2pdist}, one sees that the peaks are of the order of about a percent.  This means that the perturbations to the background are small, and such small changes cannot be observed on an event-by-event basis, but only in a large sample of events. This is why the observables are the two(or more)-particle correlation functions, in which the non-trivial correlations are separated from  the uncorrelated background. Note, that not only fluctuations in different events are uncorrelated, but also  {\em statistically independent} fluctuations at different locations in the transverse plane in the same event. 

In the two-particle correlation functions one measures mean squares of the perturbations. Therefore the smallness of the perturbation appears quadratically, and thus one has to be able to  get to the level between $10^{-3}$ and $10^{-4}$  or so in the correlation magnitude. Nevertheless,   the large set of the recorded events ($\sim 10^9$) by RHIC or LHC detector, with $\sim 10^3$ particles or  $\sim 10^6$ particle pairs per event provides a sufficient statistical data sample. 

Let us now proceed with our theoretical calculation of the two-body correlation function based on the single-particle distribution resulting from the Green function  (point-like perturbation).  These correlation functions are presented in two forms, which in fact contain  equivalent information: as a function of the relative azimuthal angle or as a ``power spectrum" of the flow harmonics. Let us start by looking at the former.

In order to calculate  the two-particle distribution one should  simply take a product of two  single-particle distributions, and perform the averaging over the random axial position of the  initial perturbation
\begin{eqnarray}
\frac{dN}{d(\Delta \phi)} \sim \int
\frac{dN}{d(\phi_1-\psi)}\frac{dN}{d(\phi_2-\psi)}d\psi
\end{eqnarray}
The averaging reduces the  function of two angles into a function
of only one, the azimuthal difference $\Delta\phi=\phi_1-\phi_2$. (This is only so for central collisions, which are axially symmetric: otherwise the situation is more complicated as the direction of the impact parameter breaks the axial symmetry. This is one of the reasons we focus on central collisions in this work.)

Our results for the two-particle distributions  for three viscosity values  are shown in the top three plots of Fig.\ref{2pdist}. Note first their distinctive shape, with a  larger peak centered at $\Delta\phi=0$ (when both particles belong to the same maximum of a single-particle distribution) and two smaller peaks at $\Delta \phi \sim \pm 2$, when two particles belong to two different peaks, connected by some flat region between them. This shape of the sound Green function is in fact very similar to what is observed experimentally, for example in the bottom plot in \ref{2pdist} which corresponds to data from ATLAS \cite{ATLAS_corr}.

Now comparing the three pictures in more detail, one observes that the upper plot (for zero viscosity) has more structure. The upper plot has four distinct ``dips" in which that two-particle distribution is less than average. Their origin is explained by matter sucked out by the passing sound front behind it.

The origin of the additional peaks next to the zero-angle one is the correlation between one of the peaks in the single-particle
distribution with matter inside the circle. These extra peaks
are attenuated when viscosity is used and for $\eta/s=0.134$ they have already disappeared. This
happens because  the viscosity induces  cancellations, between the negative ``suction regions" and positive extra
matter inside the circle. 

There are now many experimental results for the two particle correlations in central collisions such as STAR collaboration data \cite{STAR} for a centrality of $0-12\%$, data from ATLAS and ALICE in the very central region $0-1\%$ \cite{ATLAS_corr},\cite{ALICE_corr}. Now, comparing  our calculated two-particle distributions Fig.\ref{2pdist} to these  data one should be impressed by a striking similarity between their shapes, especially for the ``realistic viscosity" (the third in Fig.\ref{2pdist} ). The width of the main peak is correctly reproduced, provided the viscosity is correct. Also the ``double-hump" structure on the away side, with the correct shape of the plateau in between is found. (The peaks are a bit shifted, it is because the sound velocity  as well as the shape/size of the freezeout surface  is not quite realistic in our analytic approach.)

Let us emphasize that this non-trivial shape  comes from the hydrodynamical calculation itself, with the initial condition simply being a local(delta function like) ``hot spot". This agreement of the shape allows us to conclude, that the experiments in question {\em do see the sound waves propagation}, by a distance comparable to the fireball radius. The angular positions of the secondary peaks depend entirely on the ratio of the ``sound horizon" to the size of the fireball (the speed of sound and the freezeout time).

All our pictures are assumed to be rapidity independent,  thus the zero-angle peak is nothing else but the so called {\em ``soft ridge"} discussed in literature as a separate phenomenon.
We are pleased to see that its height, with respect to the two other peaks, is about the same as in the data, especially for the third case in Fig.\ref{2pdist} ). The angular width of this main peak is, in this case, also quite close to the data.

\subsection{The power spectrum and the initial width of the perturbation}
We have also calculated the so called ``power spectra" for the two-particle correlation functions.  Those either can be calculated from the Fourier transform of the correlator as a function of $\Delta \phi$, $C_n$, or from  the modulus squared of the flow harmonics in the single particle spectrum, since $C_n=\textrm{v}_n^2$. In this last form the expansion of the two-particle correlation function is 
 \be {dN \over d\Delta \phi}= 1+ 2 \sum_m |\textrm{v}_m|^2 cos(m\Delta \phi).
\ee 
and thus it carries the same information as the power spectrum of harmonics, in which $ |\textrm{v}_m|^2$ are plotted versus m.
(Notice that these v$_m$ are the coefficients of the Fourier expansion of the particle distribution and are not to be confused with the velocity coefficients $v_l(\rho)$ of the perturbation). The main advantage of studying the  power spectrum is that the phenomena associated with higher harmonics becomes  more visible,
which is difficult to see in the correlation function itself.

\begin{figure}[ht!]
\begin{center}
\includegraphics[width=8. cm]{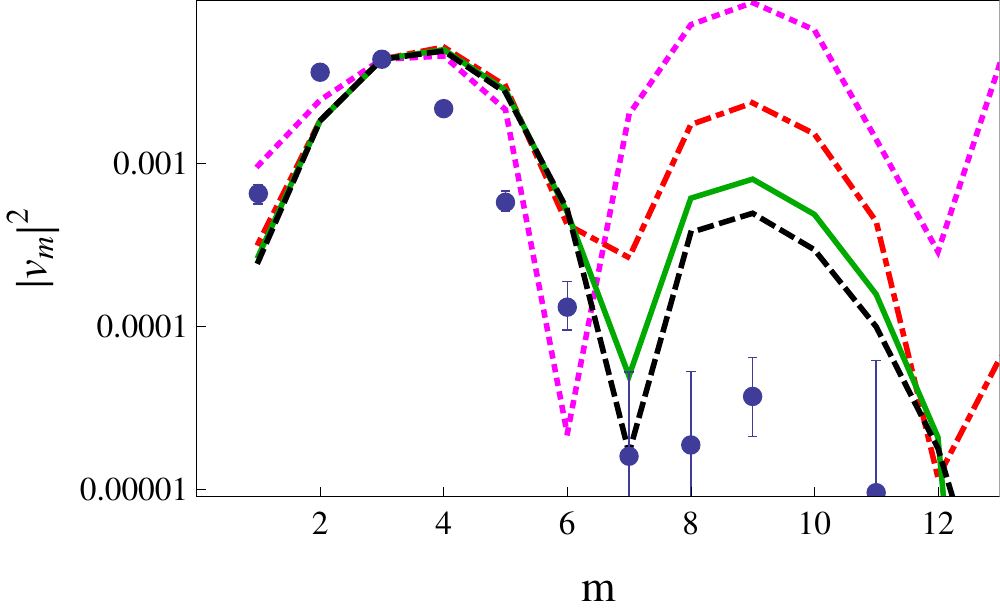}
\includegraphics[width=8. cm]{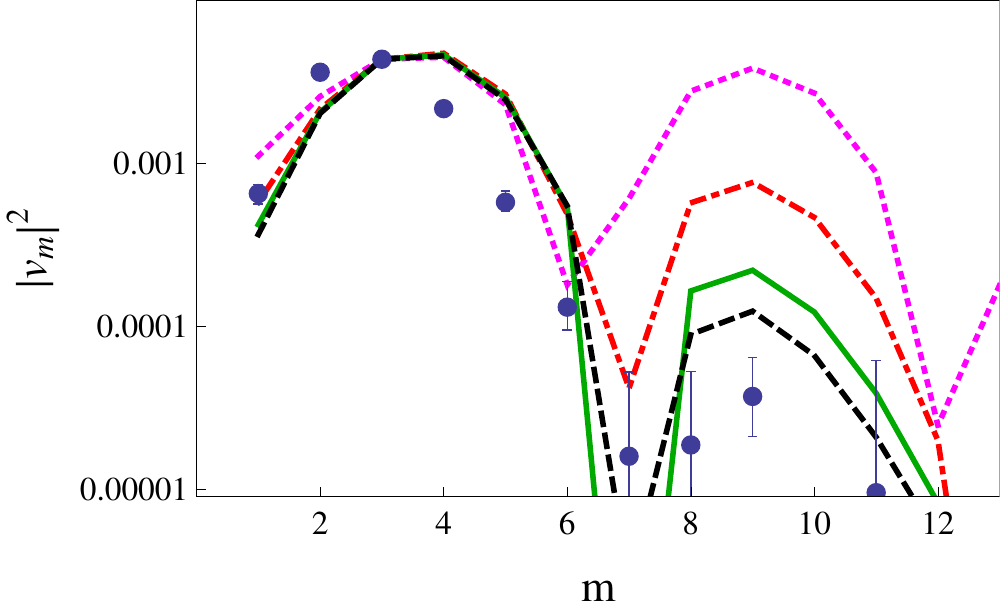}
\includegraphics[width=8. cm]{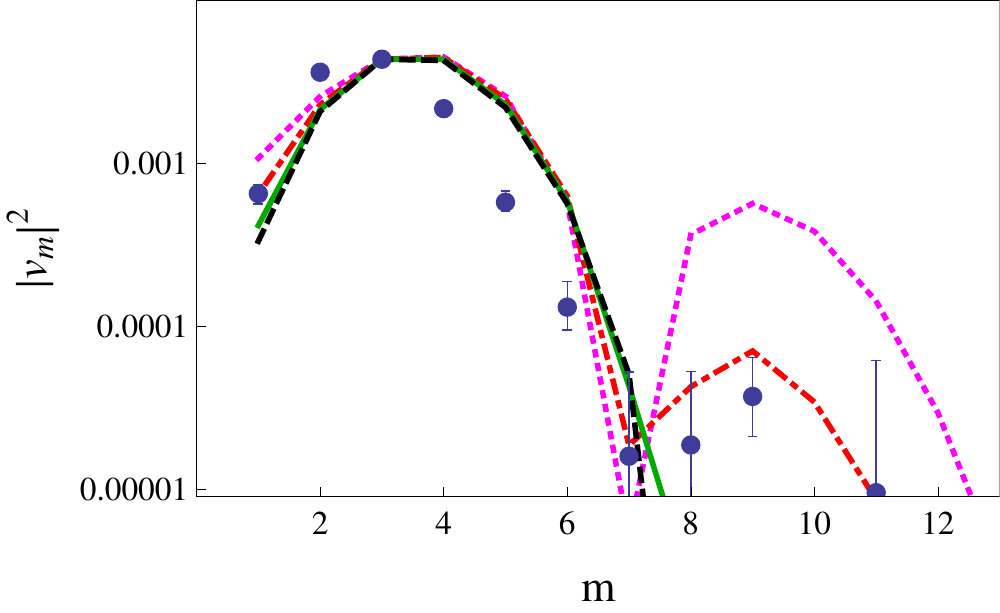}
\end{center}
\vspace{-5ex}\caption{(Color online) Spectral plots for three for three widths of the initial perturbation, 0.4,0.7 and 1 fm, from top to bottom.
The (magenta) small-dashed, the (red) dash-dotted, the (green) solid and (black) dashed curves are for $\eta/s=0,0.08,0.134,0.16$,
respectively. The data points are preliminary data  from ATLAS reported at QM2001 \cite{ATLAS_corr}. Similar data 
(not shown here) have been reported by the PHENIX  \cite{PHENIX_corr} and STAR \cite{STAR_QM} collaborations.
All the curves are arbitrarily normalized to fit the third harmonic.
}\label{fig_spectral}
\end{figure}

The result is shown in Fig.\ref{fig_spectral} and it presents  maxima and minima. 
This structure of the power spectrum, with several ``acoustic peaks", is known also for other  oscillations, most notably for those seen in the power spectrum of the angular harmonics of the Cosmic Microwave Background (CMB) distribution over the sky such as the famous Fig.9 of \cite{7WMAP}. 
Both in the Big and Little Bangs, the time allocated to the hydrodynamical stage of the evolution is limited by the so called ``freezeout time" $\tau_f$, after which the collision rates in  matter can no longer keep up with the system's expansion. At this time the propagation of the elastic waves stops and each harmonic has  at this moment  a different  phase of its oscillation.
  
   While the CMB measurements read the temperature perturbation $\delta(fo)$ directly from the sky, and thus the nodes of  $\delta_l(fo)$ correspond to the minima, in  the Little Bang one has to calculate the specific combination of the temperature and flow perturbations. This includes the calculation of how the freezeout surface is modified, which was done in  preceding sections. It is the nodes/maxima of this ``observable" combination which make the acoustic minima/maxima.   
   Note that the simple physics behind this argument makes it very robust. The minima/maxima are easily predictable and rather insensitive to many details such as dissipation. 
In fact the only assumption needed for this idea to be used in practice
is that the initial state perturbations $\delta_l(in)$ do $not$ have an oscillatory dependence on $l$ of their own.

  Before we discuss the results, we need to mention another important parameter of the problem, namely the $size$ of the initial perturbation.
In all the discussion above this was taken as small and thus unimportant: one could think of the perturbation as being practically point-like,
and thus the results being basically the Green function of the equations we are solving. However, as we will see shortly, when one discusses   
the magnitude of the higher harmonics, this size does matter. 

Fig.\ref{fig_spectral} shows how this  works in practice, the three plots correspond to three different widths of the initial perturbation: 0.4,0.7 and 1 fm,  and as one can see a change in this size does change significantly the tail of higher harmonics, the larger the width the smaller the height of the larger harmonics in the power spectrum.
Nevertheless, this does not affect the location of the acoustics dip and the secondary maximum, which remain around $m=7$ and 9, respectively.

Different curves on the plot correspond to different viscosities (see the caption), and as one can see, they do affect higher harmonics drastically.
This is to be expected, as higher harmonics of the flow have higher gradients of the flow. 
One can also see from these figures that the fit to the viscosity value must be done together with the fit to the initial size, as they are very much correlated with each other. 

We will not attempt an actual fit here, adding just some comments about the issues encountered.
The physics of the initial perturbation size  should be, first of all,  related to the size of the ``gluonic spot" in a nucleon, propagated via pQCD
evolution to appropriate $x$ and scale $ Q$ under consideration. At RHIC, with $x\sim 10^{-2},Q\sim 1-2\, GeV$  we know from DESY 
experiments (e.g. diffractive $J/\psi$ production) it to be rather small, of about $.3 \,fm$. But then there is some non-equilibrium 
stage, before hydro equations become valid, during which this spot should grow. To define the particular value one needs to know
the non-equilibrium physics at this stage. Even to define the start of hydro, one needs to know which version of hydro is used, ideal, viscous or
``resummed": for recent discussion of these issues refer to \cite{LS} and references therein.  
  One more comment on the plots in Fig.\ref{fig_spectral} is perhaps in order: as the reader can see, the curves look shifted toward the larger
  $m$ from the data points, especially well seen for $m=4..6$. Larger $m$ corresponds to smaller angular size of the sound circles. 
  This happens because we have not fitted the freezeout temperature and  time $\tau_f$ to these data: decreasing the former and increasing
  the latter one can certainly get better fit. We have not done so because in any case our calculation is done for conformal matter with fixed
  speed of sound and $\epsilon/T^4$, and cannot accurately describe the real collisions anyway.  

\subsection{The location of the perturbation}

   So far we have demonstrated some qualitative features of the one-body spectrum and two-body correlations resulting from
   a local perturbation, selecting one typical location. In this section we provide further detail on 
   the  modifications of the  Green function we calculated on the location of the initial hot spot.  
\begin{figure}[!h]
\begin{center}
\includegraphics[width=7 cm]{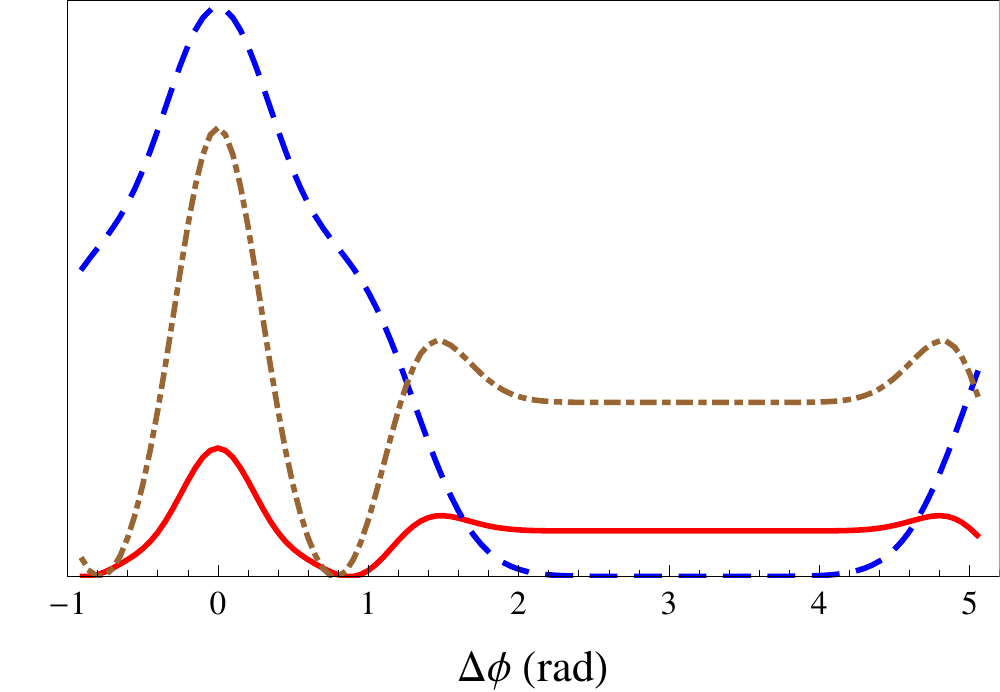}
\includegraphics[width=7 cm]{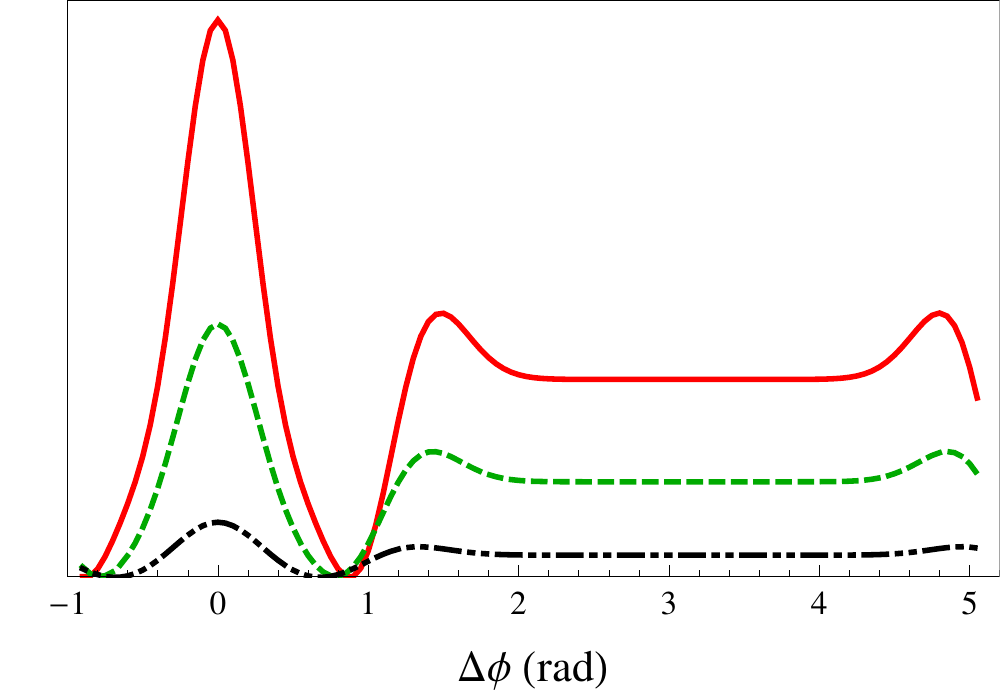}
\end{center}
\vspace{-5ex}\caption{(Color online) Top: The two-pion
distribution in arbitrary units as a function of azimuthal angle difference
$\Delta\phi$ (rad),  for $r=$2(blue large
dash),3(brown dash-dot),4.1(red solid line) fm. Bottom: The two-pion
distribution in arbitrary units as a function of azimuthal angle difference
$\Delta\phi$ (rad),  for $r=$4.1(the same red solid line),4.7
(green small dash),5.5 (black dash-dot-dot) fm. All plots are for
the same value of the viscosity-to-entropy ratio 
$\eta/s=0.134$}\label{2pdist_location}
\end{figure}   
 Since we only consider central collisions, by ``location" we mean the radial position of the ``hot spot". As shown in Fig.\ref{2pdist_location}, 
 changing the location of the spot 
  visibly affects the quantitative shape of the two-particle correlation as well as the power spectrum  Fig.\ref{spectral_rdist}.
\begin{figure}[!h]
\begin{center}
\includegraphics[width=7.5 cm]{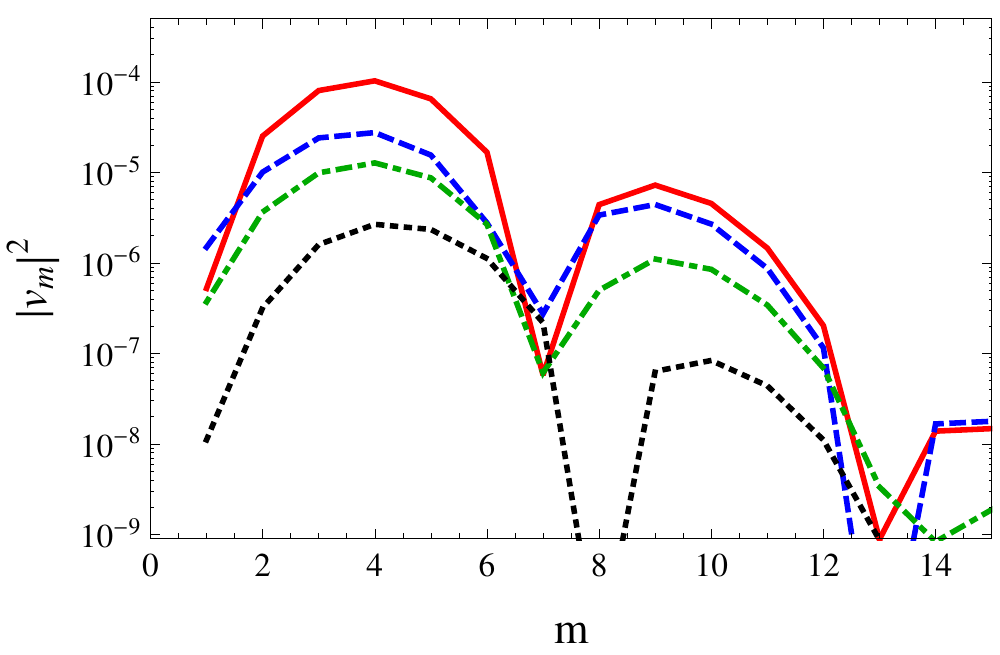}
\end{center}
\vspace{-5ex}\caption{(Color online) The(red) solid, (blue) dashed, (green) dash-dotted and (black) dotted curves correspond spectral distributions obtained for initial perturbations located at $r=3, 4.1, 4.7$ and 5.5 fm, respectively, for $\eta/s=0.134$. }
\label{spectral_rdist}
\end{figure}
When the spot is located near the center of the fireball, the two particle correlation presents only one peak located at $\Delta \phi=0$, and no structure on the away side. 
The characteristic two peaks appear when the initial perturbation is located not too close to the center($r \sim 3-5 \, fm$).  

Furthermore, as one can see, the amplitude of the modulation decreases in this case. This happens not because of a change of the hot spot amplitude (which is the same in all cases), but because of the (partial) cancellation between hydro perturbations for velocities of the first type (in the sound wave) and the second type (extra radial flow stemming
from the modification of the freezeout surface). 
As we have discovered, the very sign of the projection of the former on the radial direction depends on the initial position of the perturbation. For perturbations located near the center of the fireball it  is positive, but as the ``hot spot" gets located at larger $r$, it
decreases becoming negative  till it gets as large as the second one  and cancels it, when the ``hot spot" is located at the very edge of the fireball.

In  Fig.\ref{spectral_rdist} it is possible to see how the change in the radial position of the initial perturbation affects the power spectrum.  Its general features remain unaltered, presenting maxima and minima in all cases, which  decrease for larger values of $m$ due to  viscosity.
 The figure shows that there is some shift with $r$ in  the position of the maxima and minima.  
 
 In order to compare our results with the experimental data, it would be necessary to average over different initial perturbations, using probability distributions for their locations and amplitudes. 
 Since the minima for the different locations do not precisely match, in an averaged case a minimum would still be present, but it would not be as pronounced as in the case of an individual initial perturbation, the whole shape of the power spectrum would be smoother, with no sharp dips.
In principle, very precise data can potentially be used to infer some information about the perturbation distribution in $r$.
Such averaging is deferred to the subsequent works, as it would require
 a particular model for the initial state. It can be the Glauber model (we discussed in our previous paper)
 or some models including the saturation phenomenon. 
\section{Summary and final comment}

By calling this work  ``the second act of hydrodynamics" we emphasize the huge progress made in the field.
From measuring the mean velocity of matter and the mean ellipticity a decade ago, the first evidences for collective flow,
we now have data providing up to the 9-th harmonics of it. With many theory results, some of them in this work, 
we also now have an understanding of how perturbations behave as $m$ grows. In short, the answer is that they are acoustic oscillations,
with certain $m$-dependent oscillation frequencies and dampings. We have found that, like in the Big Bang, rotating phases at the freezeout generate minima and maxima. Remarkably, experimental data provide the first hints for the minimum and the second maximum.
   
The rather intricate shape of the two-particle correlations  as a function of $\Delta \phi$ is  very similar to the results of our calculation of the  Green function from a local source. But we would like to mention, as a parting comment, that  the questions: Do the sound circles exist in reality, or is it just a  mathematical tool ?
Are different harmonics coherent or not?  are still unanswered an they represent the next challenge for the field. 
A way to figure this out is explained in our previous paper \cite{Staig:2010pn}: one should measure  the $three$-particle
correlation functions, and look for the ``resonances" between 3 harmonics related by the ``triangular" condition
$m_1+m_2+m_3=0$, or by the two-particle correlations with
respect to reaction plane (for non-central collisions).

\vskip .25cm {\bf Acknowledgments.} \vskip .2cm The work of ES is
supported in parts by the US-DOE grant DE-FG-88ER40388, and PS is
supported by a Fulbright-CONICYT fellowship.  Helpful discussion
with S.Gubser and D.Teaney, who had  shared some of their results prior to
publication, are greatly acknowledged.


\begin{thebibliography}{99}
\bibitem{comment1} Mocsy and Sorensen had also made a popular web page, The sound of the ``Little Bang", in which they suggest
this sound to be sinusoidal with certain frequency. As is clear from the present paper, it is not like this, being instead
a single short pulse, like the boom from a passing supersonic jet.
\end{thebibliography}
\end{document}